\documentclass{ws-mpla}
\usepackage[super]{cite}
\usepackage{graphicx}
\usepackage{pdflscape}
\usepackage{longtable}
\usepackage{lipsum} 
\usepackage[colorlinks,citecolor=blue,linkcolor=red,anchorcolor=blue,filecolor=blue,urlcolor=blue]{hyperref}
\begin{document}

\markboth{N. Biswal, Nishu Jain, Raj Kumar, A. S. Pradeep, and M. Bhuyan}
{Structural and decay properties of superheavy nuclei}

\catchline{}{}{}{}{}

\title{Structural and decay properties of nuclei appearing in the $\alpha$-decay chains of $^{296,298,300,302,304}$120 within the relativistic mean field formalism}

\author{N. Biswal}
\address{Department of Physics, North Odisha University, Sri Ram Chandra Vihar 757003, India }
\author{Nishu Jain, Raj Kumar}
\address{School of Physics and Materials Science, Thapar Institute of Engineering and Technology, Patiala 147004, India \\
rajkumar@thapar.edu}
\author{A. S. Pradeep}
\address{Department of Physics, Rajiv Gandhi University of Knowledge Technologies, Andhra Pradesh 516330, India }
\author{M. Bhuyan}
\address{Center for Theoretical and Computational Physics, Department of Physics, Faculty of Science, University of Malaya, Kuala Lumpur 50603, Malaysia \\
Atomic Molecular and Optical Research Group, Advanced Institute of Materials Science, Ton Duc Thang University, Ho Chi Minh City, Vietnam \\
Faculty of Applied Sciences, Ton Duc Thang University, Ho Chi Minh City, Vietnam \\
mrutunjaya.bhuyan@tdtu.edu.vn}

\maketitle

\pub{Received (Day Month Year)}{Revised (Day Month Year)}

\begin{abstract}
\noindent
An extensive study of $\alpha$-decay half-lives for various decay chains of isotopes of $Z$ = 120 is performed within the axially deformed relativistic mean-field (RMF) formalism by employing the NL3, NL3$^*$, and DD-ME2 parameter set. The structural properties of the nuclei appearing in the decay chains are explored. The binding energy, quadrupole deformation parameter, root-mean-square charge radius, and pairing energy are calculated for the even-even isotopes of $Z$ = 100 $-$ 120, which are produced in five different $\alpha$-decay chains, namely, $^{296}$120 $\rightarrow$ $^{260}$No, $^{298}$120 $\rightarrow$ $^{262}$No, $^{300}$120 $\rightarrow$ $^{264}$No, $^{302}$120 $\rightarrow$ $^{266}$No, and $^{304}$120 $\rightarrow$ $^{268}$No. A superdeformed prolate ground state is observed for the heavier nuclei, and gradually the deformation decreases towards the lighter nuclei in the considered decay chains. The RMF results are compared with various theoretical predictions and experimental data. The $\alpha$-decay energies are calculated for each decay chain. To determine the relative numerical dependency of the half-life for a specific $\alpha$-decay energy, the decay half-lives are calculated using four different formulas, namely, Viola-Seaborg, Alex-Brown, Parkhomenko-Sobiczewski, and Royer for the above said five $\alpha$-decay chain. We notice a firm dependency of the half-life on the $\alpha$-decay formula in terms of $Q_{\alpha}$-values for all decay chains. Further, the present study also strengthens the prediction for the island of stability in terms of magic number at the superheavy valley in the laboratories.

\keywords{Relativistic Mean Field; Nuclear Structure; Decay Energy and Half-life; Isospin Asymmetry; Superheavy Nuclei}
\end{abstract}



\maketitle
\section{Introduction}
The quest for the synthesis of exotic drip-line and superheavy nuclei (SHN) attracts the great interest of physicists worldwide in the last two decades. The difficulty in synthesizing superheavy elements (SHE) is that they are highly unstable and hard to detect. With increase in the atomic number, the repulsive Coulomb potential increases compared to the attractive nuclear potential, which leads to a decrease in its stability. Furthermore, the fusion cross-section of the SHN is of the order of picobarns (pb) and also have very short half-lives (as low as milliseconds).  Meanwhile, the shell correction energy plays a significant role in stabilizing the nucleus against the Coulomb repulsion.  The estimations of the structural properties of the superheavy nuclei are motivated by the “island of stability” at or near isospin asymmetry, beyond the spherical doubly-magic nucleus $^{208}$Pb and carry profound importance in contemporary nuclear physics  \cite{Dong12,Giu19,Naza18,bhu12,zhang05,meng96,meng98,meng06}. The latest superheavy nuclei added to the periodic table is $Z$ = 118 with the name Oganessian (Og), which was synthesized in hot fusion at JINR, Dubna \cite{Oga07} followed by preceding elements at GSI, Germany, and RIKEN, Japan via cold fusion \cite{hofmann2000,hofmann11,morita07}. Now the interest has been shifted to the synthesis of superheavy nuclei with $Z$=120 \cite{ogan09}, as this may be the next candidate for the magic island for extra-stable nuclei in the sea of fission-instability, as predicted by various theoretical models \cite{bhu12,rutz97,bend99,krup00,rein02,bend01,meng96,meng98,zhang05}. The efforts to synthesize new elements via hot or cold fusion reactions and examine their structural properties have been going on with the support of self-consistent microscopic theoretical models \cite{bhu12,zhang05,Dvora06,Barb11,karol16,kumarprcr2020,liang15,shen19,ren20}. 

The decay properties of superheavy nuclei are strongly connected with the stability via shell effects, isospin dependency of the nuclear structure, nuclear deformation, rotational and vibrational properties, occupation and/or single-particle energy level, and fusion-fission dynamics \cite{nils69,fise72,stasz13,bhu15,bhu18,bhu11a,bhu11,bhu09,kumarprc2015}. Most of the proton-rich SHN are identified experimentally via $\alpha$-decay chains, which provides important information about the influence of nuclear structure on the decay properties of these nuclei. A superheavy nucleus predominantly undergoes sequential $\alpha$-decay (s), followed by subsequent spontaneous fission. An $\alpha$-decay bears a great chance to test different nuclear phenomena in finite nuclei and an astrophysical process. The $\alpha$-decay properties of the new $\alpha$-emitters could be correlated with those of previously known $\alpha$-emitters, and the systematic trends that existed could be employed in predicting the properties of $\alpha$-emitters \cite{perl50}. A lot of theoretical studies have been performed to predict $\alpha$-decay half-lives \cite{pei07,kelk07,chow08,roye08,mohrprcr17,santhoshprc18,denisovprc10,srid19,heen15,wang13}. At present, a force-independent theoretical framework is required to predict the ground state bulk and decay properties of the superheavy nuclei. In last few decades, theoretical frameworks based on the nuclear mean-field approach have been successfully applied to nuclear structure studies and classified into two categories. On the one hand, the microscopic-macroscopic (mic-mac) models, including a liquid-drop picture of the nucleus, where the nuclear part of the energy varies smoothly with the numbers of nucleons and a shell correction contribution obtained from a suitable single-particle potential for the fine structure  \cite{sharma05,Strut67,Stru68}. On the other hand, the microscopic Hartree-Fock and/or Hartree calculations based on Skyrme forces and/or the non-linear relativistic mean-field model \cite{bhu12,bhu09,bhu11,bhu11a,bhu15,bhu18,chab97,ston07,bend03,sil04,bhu19,zhang05,meng98,wang13} applied successfully to obtain the ground and intrinsic excited state self-consistently. The detail of these various Skyrme and relativistic mean-field parameter sets are given in the Refs. \cite{dutra12,dutra14}.

At a microscopic level, the non-relativistic and relativistic mean-field models have received wide attention due to their successful description of a large variety of nuclear phenomena during the past three decades and are viewed as a standout amongst other microscopic candidates for the descriptions of both exotic and superheavy nuclei \cite{chab97,zhang05,bhu12,ston07,bend03,sil04,meng06,shar93,ring90,ring96,sero86,bhu09,bhu11,bhu15,bhu18,bhu18a,bhu19}. The Skyrme-Hartree-Fock and the relativistic mean-field model with various conventional parameter sets predict $Z$ = 120 with $N$ = 182 and/or 184 as the best candidates for spherical shell closures on the island of superheavy nuclei  \cite{rutz97,bend99,krup00,rein02,bend01,bhu12,zhang05}. Of course, the different nature of the spin-orbit interaction in the Skyrme and RMF models is pivotal in deciding the location of the more substantial shell effects over the isotopic chain of $Z$ = 120 \cite{zhang05,bhu12} and references therein. Furthermore,  the fusion studies \cite{hoff08,ogan09,kozu10,hein13,chris17,bhu20,rana20} have also been performed with various target-projectile combinations leading to different isotopes of $Z$=120 with a wide range of synthesis probabilities. In this context, it is one of the prime objectives to study various decay chains of $Z$ =120, namely, $^{296}$120 $\rightarrow$ $^{260}$No, $^{298}$120 $\rightarrow$ $^{262}$No, $^{300}$120 $\rightarrow$ $^{264}$No, $^{302}$120 $\rightarrow$ $^{266}$No, and $^{304}$120 $\rightarrow$ $^{268}$No. This includes the $\alpha$-decay energies ($Q_\alpha$) that are estimated from the binding energies of parent and daughter nuclei from the relativistic mean-field model for popular NL3 \cite{lala97}, recently developed NL3$^*$ \cite{lala09}, and density-dependent DD-ME2 \cite{niks05} parameter sets. The $Q_\alpha$s are further used to obtain the respective half-lives ($T_\alpha$) from four different well-known $\alpha$-decay formulae of Viola-Seaborg \cite{viola66,sobic1989}, Brown \cite{brown1992}, Parkhomenko-Sobiczewski \cite{menko}, and Royer-Dasgupta-Reyes \cite{royer2000,dasgupta}, to obtain the relative numerical dependence of the half-life for a specific $\alpha$-decay energy. It is worth mentioning that these formulae are known for predicting the $\alpha$-decay half-lives in the superheavy mass region. Hence, it would be quite interesting to compare the relative outcome of these analytical formulae in terms of the Q-value of the $\alpha$-decay.

This paper is structured as follows: In Sec. \ref{theory}, we briefly present the relativistic mean-field model. Sec. \ref{results} and Sec. \ref{decay} shows the relativistic mean-field calculations for structure and decay properties. Finally, in Sec. \ref{summary}, we discuss brief conclusions and perspectives.\\


\section{Relativistic mean field model}
\label{theory}
The relativistic mean-field (RMF) model is a microscopic approach to solve the many-body problem through the interacting meson fields. The model predicts the ground and the intrinsic excited-state structural property of finite nuclei throughout the nuclear landscape. The details of the RMF models and their parametrizations can be found in Ref. \cite{dutra14} and references therein. A typical form of RMF Lagrangian density is given as \cite{bogu77,sero86,ring86,lala99c,bhu09,bhu11,bhu15,bhu18,bhu18a,rein89,ring96,vret05,bret00,meng06,paar07,niks11,logo12,zhao12,lala09,lala97,niks05,meng98a,meng13,zhao18},
\begin{eqnarray}
{\cal L}&=&\overline{\psi}\{i\gamma^{\mu}\partial_{\mu}-M\}\psi +{\frac12}\partial^{\mu}\sigma
\partial_{\mu}\sigma -{\frac12}m_{\sigma}^{2}\sigma^{2} 
-{\frac13}g_{2}\sigma^{3} -{\frac14}g_{3}\sigma^{4} \nonumber \\
&& -g_{s}\overline{\psi}\psi\sigma -{\frac14}\Omega^{\mu\nu}\Omega_{\mu\nu} +{\frac12}m_{w}^{2}\omega^{\mu}\omega_{\mu}
-g_{w}\overline\psi\gamma^{\mu}\psi\omega_{\mu} -{\frac14}\vec{B}^{\mu\nu}.\vec{B}_{\mu\nu} \nonumber \\
&& +\frac{1}{2}m_{\rho}^2
\vec{\rho}^{\mu}.\vec{\rho}_{\mu} -g_{\rho}\overline{\psi}\gamma^{\mu}
\vec{\tau}\psi\cdot\vec{\rho}^{\mu} -{\frac14}F^{\mu\nu}F_{\mu\nu} -e\overline{\psi} \gamma^{\mu}\frac{\left(1-\tau_{3}\right)}{2}\psi A_{\mu}.
\label{lag}
\end{eqnarray}
with vector field tensors
\begin{eqnarray}
F^{\mu\nu} = \partial_{\mu} A_{\nu} - \partial_{\nu} A_{\mu} \nonumber \\
\Omega_{\mu\nu} = \partial_{\mu} \omega_{\nu} - \partial_{\nu} \omega_{\mu} \nonumber \\
\vec{B}^{\mu\nu} = \partial_{\mu} \vec{\rho}_{\nu} - \partial_{\nu} \vec{\rho}_{\mu}.
\end{eqnarray}
Here the fields for the $\sigma$-, $\omega$- and isovector $\rho$- meson is denoted by $\sigma$, $\omega_{\mu}$, and $\vec{\rho}_{\mu}$, respectively. The electromagnetic field is defined by $A_{\mu}$. The quantities, $\Omega^{\mu\nu}$, $\vec{B}_{\mu\nu}$, and $F^{\mu\nu}$ are the field tensors for the $\omega^{\mu}$, $\vec{\rho}_{\mu}$ and photon fields, respectively.

The density dependence of the meson-nucleon coupling in the relativistic mean-field model can be parameterized in terms of a phenomenological approach as:
\begin{eqnarray}
\centering 
\label{eqn:7}
g_{i}(\rho)=g_{i}(\rho_{sat})f_{i} (x)\vert_{i=\sigma,\omega},
\end{eqnarray}
where
\begin{eqnarray}
\centering 
\label{eqn:8}
f_{i}(x)=a_{i}\dfrac{1+b_{i}(x+d_{i})^{2}}{1+c_{i}(x+d_{i})^2}
\end{eqnarray}
and
\begin{eqnarray}
\centering 
\label{eqn:9}
g_{\rho}=g_{\rho}(\rho_{sat})e^{a_{\rho}(x-1)}
\end{eqnarray}	    
Here, $x=\rho/\rho_{sat}$ and the eight real dependent parameters can decrease to three in the number by owing the five constraints viz., $f _{i}(1)=1 $, $f _{\sigma}^{''}(1)$=$f _{\omega}^{''}(1)$ and $f _{i}^{''}(0)=0 $. More details and recent advancements in the relativistic mean-field can be found in Refs.\cite{niks05,lala97,lala09,niks11,bhu18,meng98a,meng13,zhao18} and reference therein.

From the above Lagrangian density, we obtain the field equations for the nucleons and the mesons. These equations are solved by expanding the upper and lower components of the Dirac spinors and the boson fields in an axially deformed harmonic oscillator basis for an initial deformation $\beta_{0}$. The set of coupled equations is solved numerically by a self-consistent iteration method. The center-of-mass motion energy correction is estimated by the usual harmonic oscillator formula $E_{c.m.}=\frac{3}{4}(41A^{-1/3})$. The quadrupole deformation parameter $\beta_2$ is evaluated from the resulting proton and neutron quadrupole moments, as
\begin{equation}
Q = Q_n + Q_p =\sqrt{\frac{16\pi}5} \left (\frac3{4\pi} AR^2\beta_2 \right ).
\end{equation}
The root mean square (rms) matter radius is defined as
\begin{equation}
\langle r_m^2\rangle={1\over{A}}\int\rho(r_{\perp},z) r^2d\tau,
\end{equation}
where $A$ is the mass number, and $\rho(r_{\perp},z)$ is the deformed density. The total binding energy and other observables are also obtained by using the standard relations, given in Ref. \cite{ring90}. The most popular NL3 \cite{lala97}, recently developed NL3$^*$ \cite{lala09}, and density-dependent DD-ME2 \cite{niks05} interaction parameters are used for the present analysis. It is worth mentioning that these interaction parameters are able to describe the bulk properties of the nuclei reasonably good from $\beta-$ stable region to the drip-line \cite{ring86,lala99c,bhu09,bhu11,bhu15,bhu18,bhu18a,rein89,ring96,vret05,meng06,paar07,lala97,niks05,niks11,logo12,zhao12,lala09}. To deal with the open-shell nuclei, one has to consider the pairing correlations in their ground as well as excited states \cite{karat10}. There are various methods, such as the BCS approach, the Bogoliubov transformation, and the particle number conserving methods, that have been developed to treat pairing correlation in the finite nuclei \cite{moli97,zhang11,hao12}. In principle, the Bogoliubov transformation is the widely used method to take pairing correlations for the drip-line nuclei \cite{vret05,paar07,meng06,bhu18,bhu19}. We have used the constant gap BCS approach within the NL3, $\&$ NL3$^*$ and Bogoliubov transformation within the DD-ME2 parameter sets for the present analysis \cite{vret05,niks05,lala97,lala09,bhu18,bhu09,bhu15,bret00,lala99c,zhao12,meng06,meng13,zhang05,meng98a}.

\begin{figure}
\begin{center}
\includegraphics[width=0.75\columnwidth]{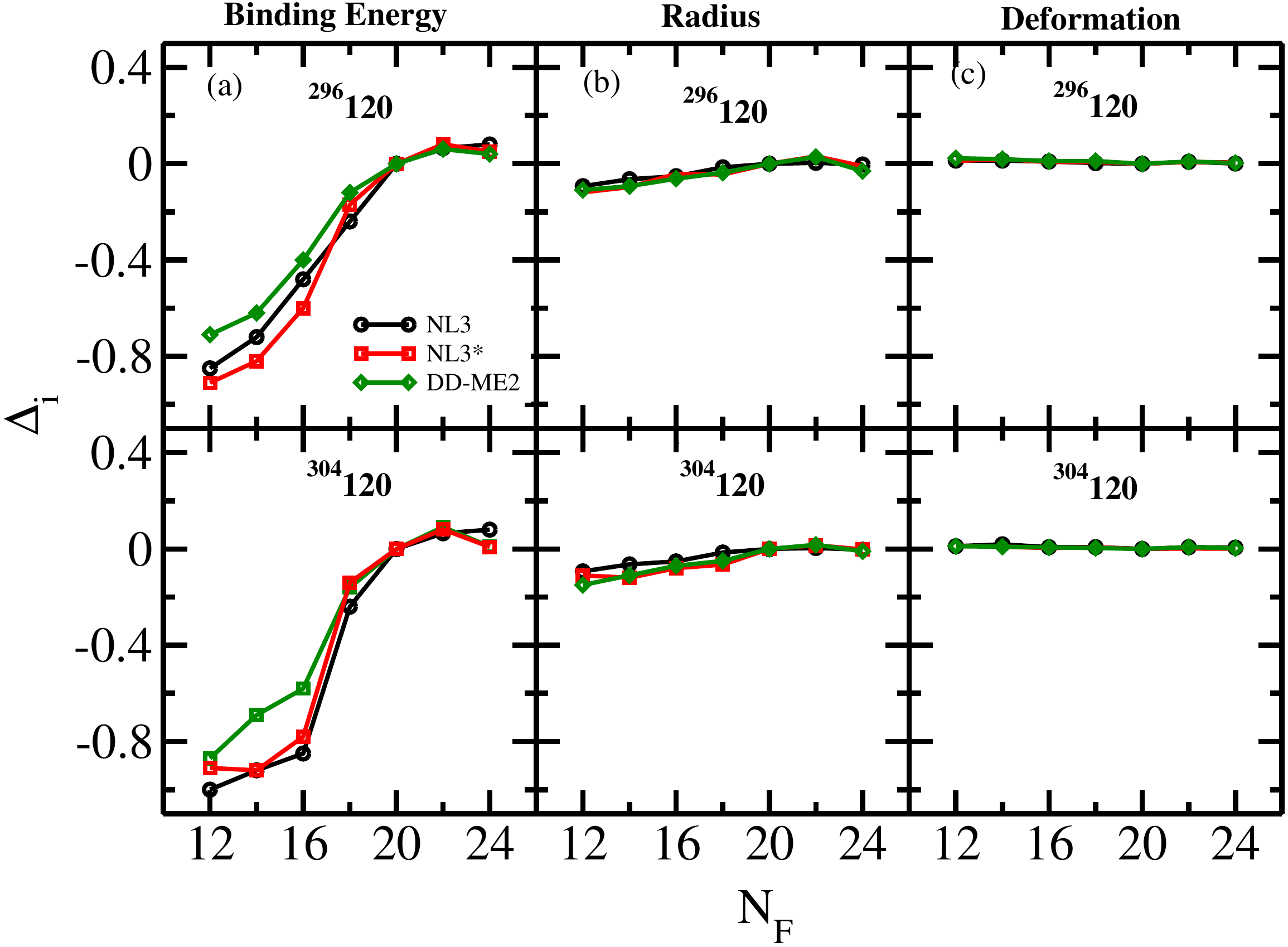}
\caption{\label{fig0a}(Color online) The relative difference in the solution, $\Delta_i$ = Solution$\vert_{N_F = 20}$ - Solution$\vert_{N_F = 12-24}$)  for (a) binding energy, (b) root-mean-square matter radius, and (c) quadrupole deformation for $^{296}$120 (upper panel), and $^{304}$120 (lower panel), respectively. The black line with an open circle, red line with an open square, and green line with a diamond symbol are for NL3, NL3$^*$, and DD-ME2 parameter set. The energy in MeV and radius in fm. See text for more details.}
\end{center}
\end{figure}
\begin{figure}
\begin{center}
\includegraphics[width=0.75\columnwidth]{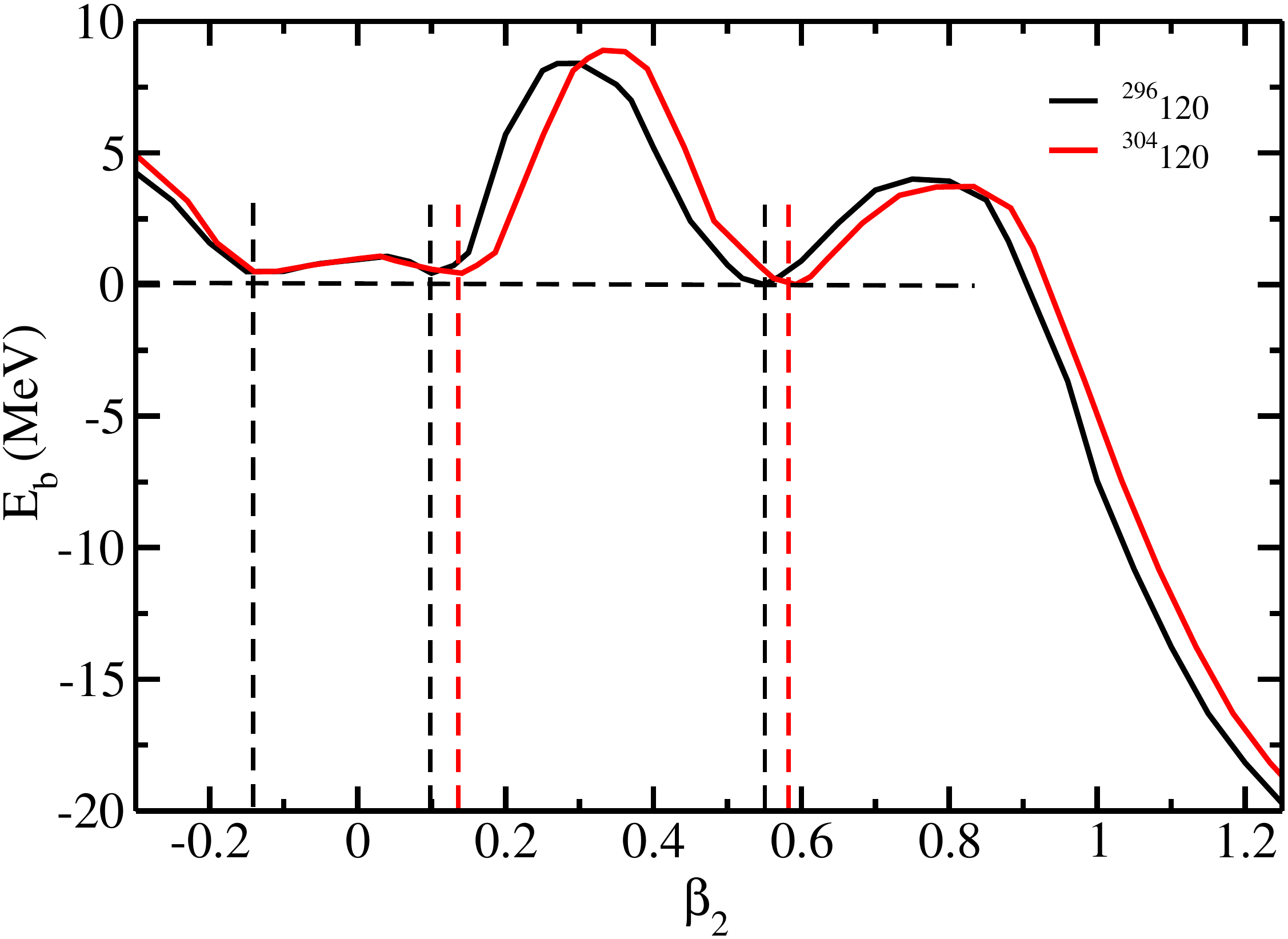}
\caption{\label{fig0b}(Color online) The potential energy surface for $^{296}$120 (black line) and $^{304}$120 (red line) as a function of quadrupole deformation parameter $\beta_2$ using the axially deformed relativistic mean approach for NL3$^*$ parameter set. See text for more details.}
\end{center}
\end{figure}

\section{Nuclear Bulk Properties}
\label{results} \noindent
The relativistic mean-field equations are solved self-consistently by taking different inputs of the initial deformation $\beta_0$ \cite{ring86,ring90,bhu09}. To verify the convergence of the ground state solutions, calculations are done for fermion shells N$_F$ = 12 - 24. The relative difference in the solutions, $\Delta_i$ = Solution$\vert_{N_F = 20}$ - Solution$\vert_{N_F = 12-24}$, where $i$ stands for binding energy, root-mean-square matter radius, and quadrupole deformation. Here for all the solutions, boson shells N$_B$ = 20. The estimated relative difference in the (a) binding energy, (b) root-mean-square matter radius, and (c) quadrupole deformation for $^{296}$120, and $^{304}$120 are shown in the upper and lower panel of Fig. \ref{fig0a}, respectively. The black line (open circle), red line (open square), green line (open diamond) are for NL3, NL3$^*$, and DD-ME2 parameter set, respectively. One can notice the relative variation of these solutions is $\leq$ 0.06$\%$ on binding energy, 0.01$\%$ on a nuclear radius, and 0.008$\%$ in the quadrupole deformation for these nuclei. From the figure, we can conclude that the desired number of major shells for fermions and bosons is N$_F$ = N$_B$ = 20. The number of mesh points for Gauss-Hermite and Gauss-Laguerre integration is 20 and 24, respectively. Here, we have employed widely used non-linear NL3, recently developed NL3$^*$, and density-dependent DD-ME2 parameter sets for the present analysis. It is worth mentioning that the NL3$^*$ \cite{lala09}, and density-dependent DD-ME2 \cite{niks05} parameter sets are then refitted, and medium dependent nucleon-meson coupling version of the NL3 \cite{lala97,lala99c}, where the exotic nature of drip-line are considered. Hence, it will be interesting to examine the applicability of relativistic mean-field for NL3$^*$ and DD-ME2 compared to the NL3 parameter set and analyze their efficiency in predicting the bulk properties of superheavy nuclei. Further, the study also provides the relative dependency of the force parameter in the structural properties of superheavy nuclei. \\

\begin{figure}
\begin{center}
\includegraphics[width=0.8\columnwidth]{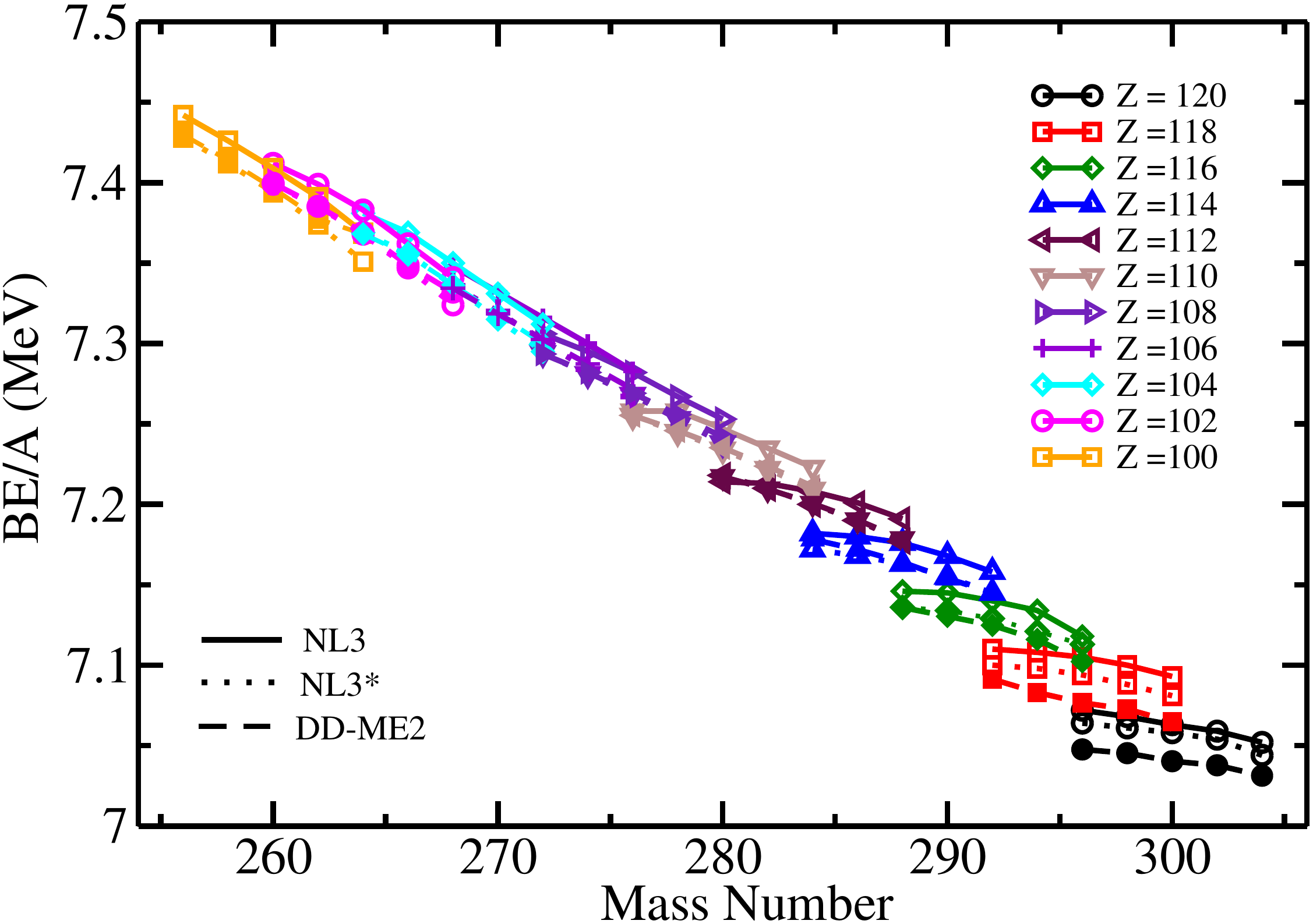}
\caption{\label{fig1}(Color online) The binding energy per nucleon for the isotopic chain of $Z$ = 100 to 120 nuclei for relativistic mean-field approach with NL3 (solid line), NL3$^*$ (dotted line), and DD-ME2 (dashed line) parameter set.}
\end{center}
\end{figure}
\begin{figure}
\begin{center}
\includegraphics[width=0.8\columnwidth]{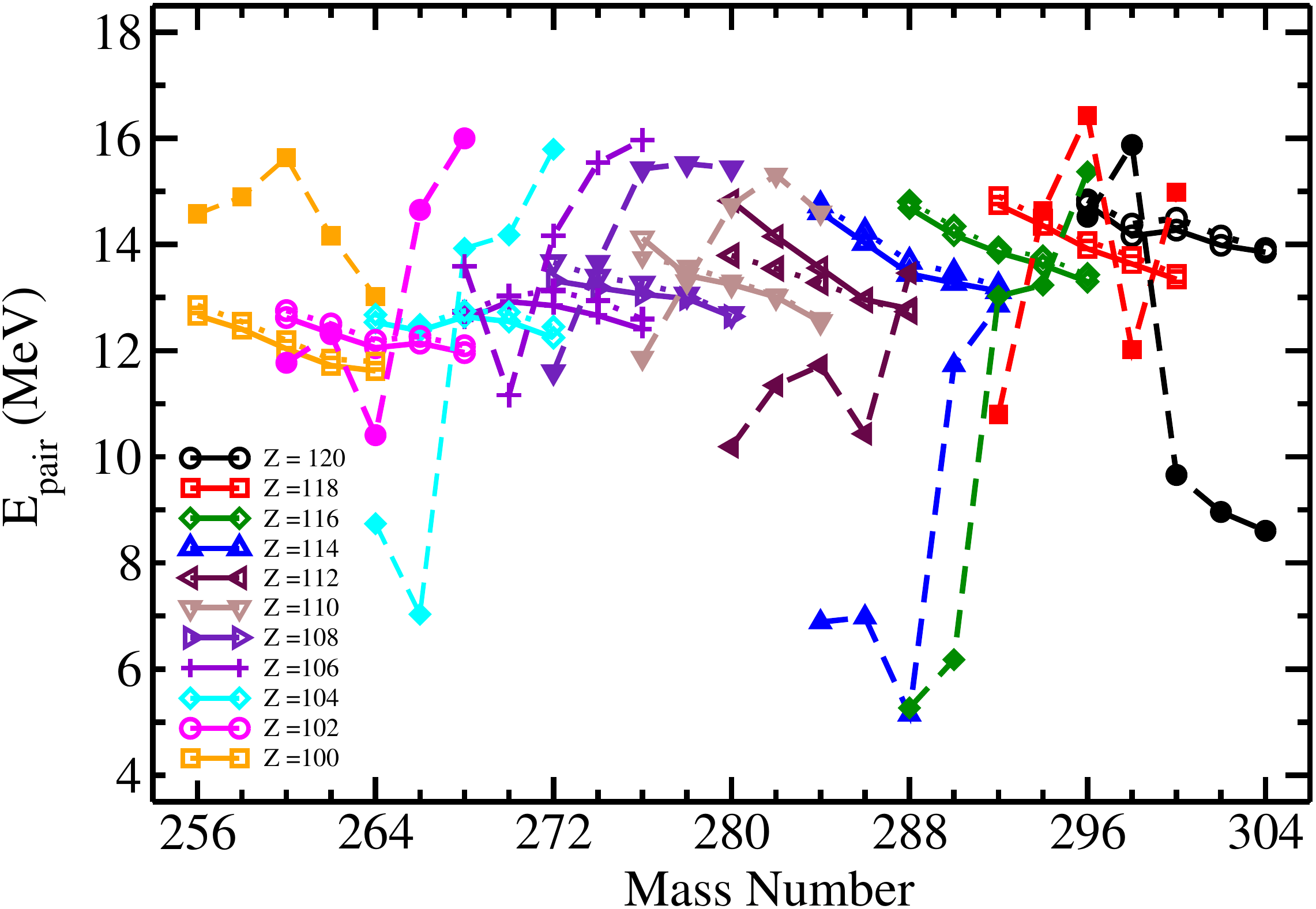}
\includegraphics[width=0.8\columnwidth]{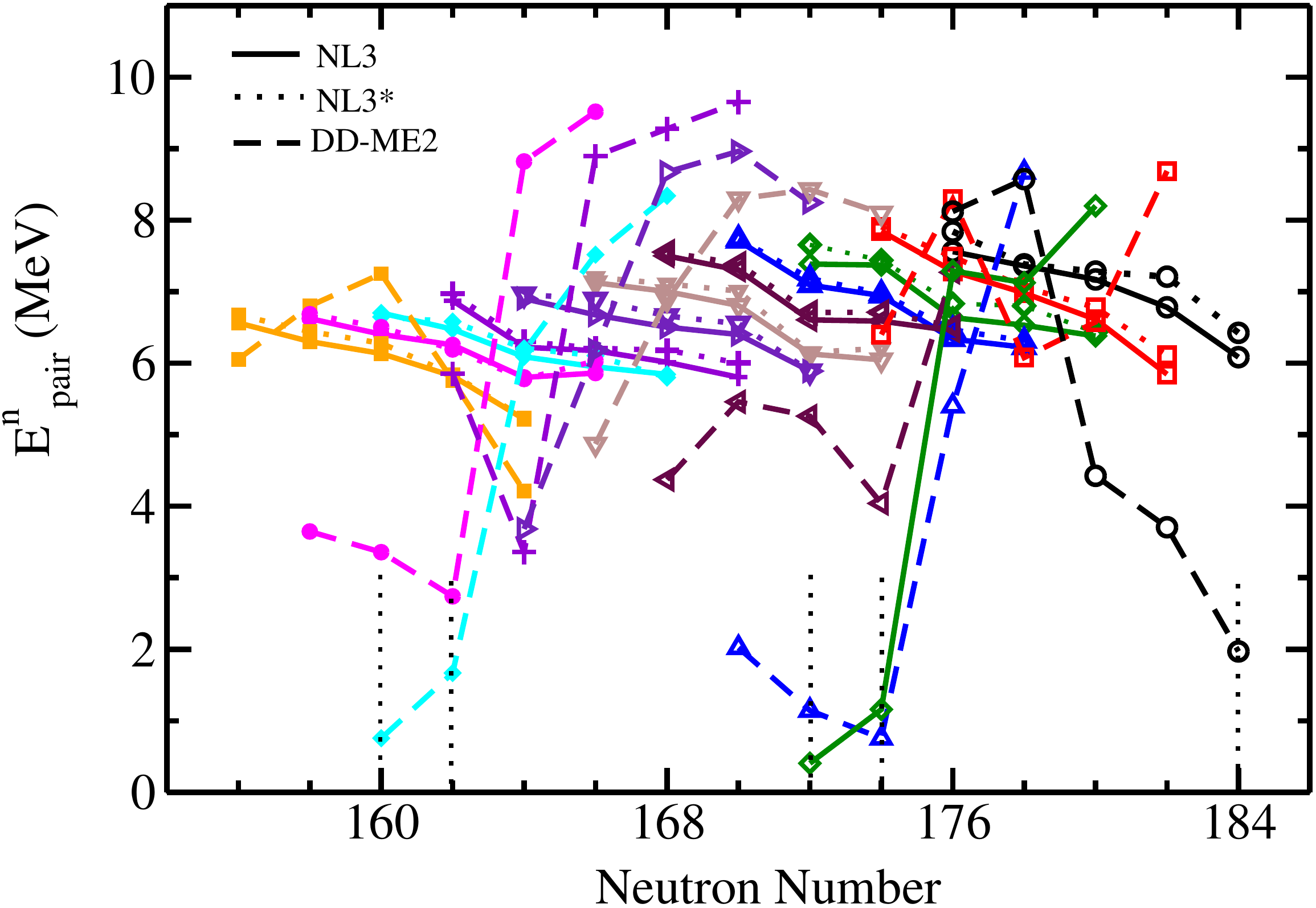}
\caption{\label{fig2}(Color online) The total $E_{pair}$ (Upper Panel) and neutron pairing energy $E^n_{pair}$ (Lower Panel) for the isotopic chain of $Z$ = 100 to 120 nuclei within the relativistic mean-field approach are for NL3 (solid line), NL3$^*$ (dotted line), and DD-ME2 (dashed line) parameter set.}
\end{center}
\end{figure}
\begin{figure}
\begin{center}
\includegraphics[width=0.8\columnwidth]{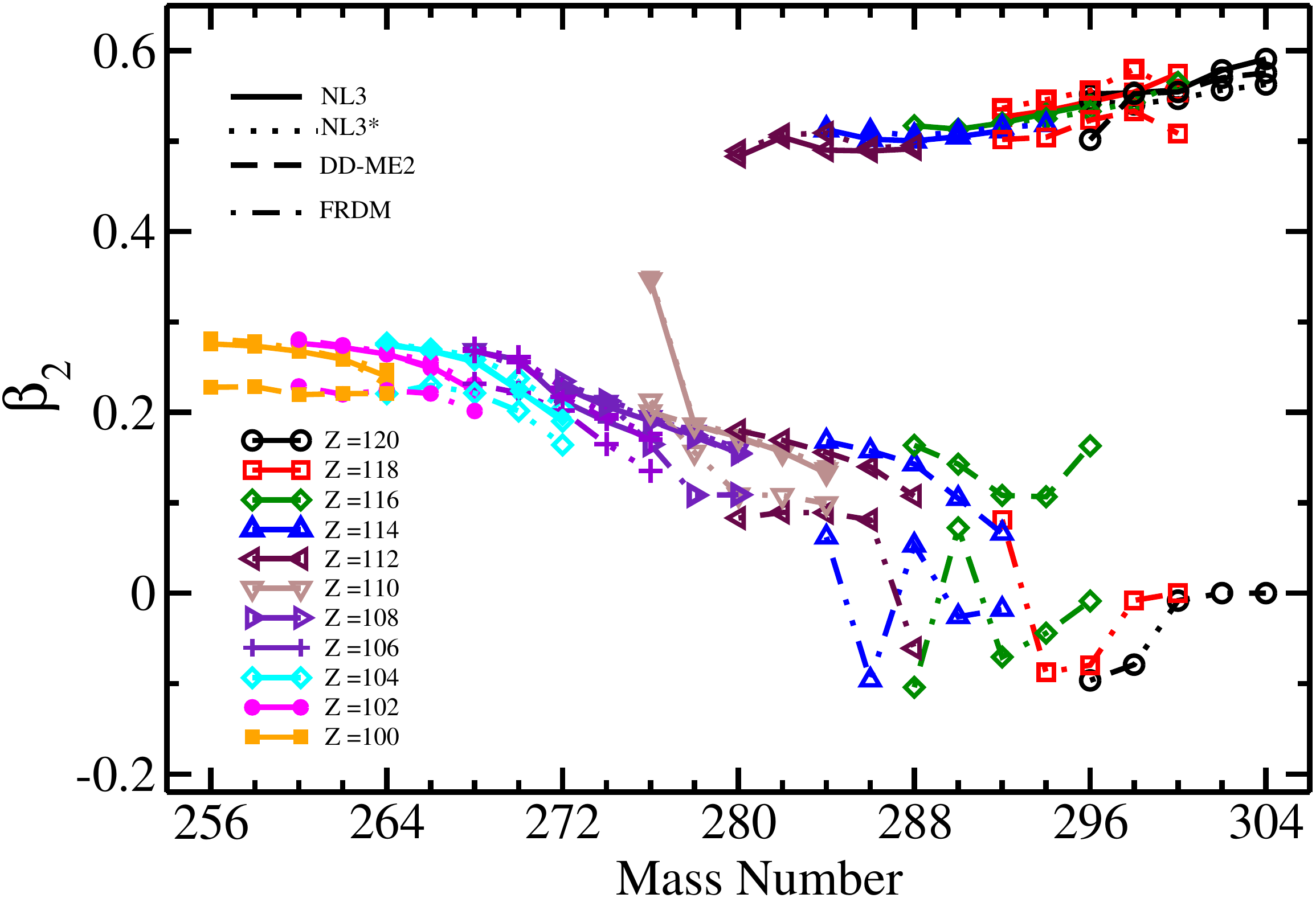}
\caption{\label{fig3} (Color online) The ground state quadrupole deformation parameter ($\beta_2$) for the isotopic chain of $Z$ = 100 to 120 nuclei within relativistic mean-field approach for NL3 (solid line), NL3$^*$ (dotted line), and DD-ME2 (dashed line) parameter set. The FRDM predictions \cite{frdm97,frdm16} (dashed line) are given for comparison.}
\end{center}
\end{figure}
\subsection{Potential energy surface}
\noindent 
Conventionally, in the case of a quantum mechanical system, the path followed by the different solutions at various deformation defines a potential barrier, which can be used to determine the ground state of a nucleus. More elaborately, the likely results for the ground state can be determined from the potential energy surface (PES) obtained from the self-consistent relativistic mean-field approach. Here, the potential energy curve is calculated microscopically using the constrained RMF theory \cite{bhu09,bhu11,lala09,ring90,bhu20}. The expectation value of the Hamiltonian at certain deformation is given as, 
\begin{equation}
H'=\sum_{ij}\frac{<\psi_i|H_0-\lambda Q_2|\psi_j>}{<\psi_i|\psi_j>}.
\end{equation} 
Here $\lambda$ is the constraint multiplier, and $H_0$ is the Dirac mean-field Hamiltonian. The convergence of the numerical solutions on the binding energy and the deformation is not very sensitive to the deformation parameter $\beta_0$ of the harmonic oscillator basis for the considered range due to the large basis. Thus the deformation parameter $\beta_0$ of the harmonic oscillator basis is chosen near the expected deformation to obtain high accuracy and low computation time cost.

The potential energy surface (PES) as a function of deformation parameter $\beta_2$ are shown for the nucleus, namely, $^{296}$120 (black line), and $^{304}$120 (red line) in Fig. \ref{fig0b}, for a representative case. The RMF (NL3$^*$) parameter is used to obtain these results. All other isotopes for $Z$ =120 are also showing respective PES curves following their ground state solutions for NL3 and DD-ME2 parameter set, which are not given here. The energy ($E_b$ = $E_{g.s}-E_{e.s}$) on the $Y-$axis is the difference between the ground state energy to other constrained solutions. The calculated PES for both the cases is shown for a wide range of oblate to prolate deformations. We notice from the figure that a few minima appear for the considered range of $\beta_2$. The magnitude of binding energy for the corresponding minima shows that the ground state solution appears at a certain value of $\beta_2$. The $\beta_2$ for the ground state is not same for all nuclei (see Table \ref{tab1}) \cite{ring90,bhu09,bhu15,bhu18,bhu20}. For example, the ground state solutions for $^{296}$120, and $^{304}$120  are $\sim$ 0.55 and 0.57, respectively, which reflect from Fig. \ref{fig0b} for NL3$^*$ parameter set. Hence, one can conclude that the ground state properties of these nuclei are independent or weakly dependent on the force parameters used.

\noindent 
\subsection{Binding Energy and Pairing Energy}
The present study explains the nuclear structure and decay properties, which deals with the basic ingredients such as binding energy (BE), quadrupole deformation parameter $\beta_{2}$, root-mean-square (rms) radius, and pairing energy, which are calculated using the relativistic mean-field approach.  We have considered the even-even isotopes of $Z$ = 98 $-$ 120, which involved the five different decay chains of $Z$ =120. Further, the analysis also demonstrates the applicability and parameter dependency of RMF for studying the nuclear structure of the superheavy nuclei. The calculated binding energy for the $\alpha$-decay chains of $^{296,298,300,302,304}$120 is obtained using axially deformed RMF model for non-linear NL3 $\&$ NL3$^*$, and density-dependent DD-ME2 parameter set. The calculated values are given along with the available finite range droplet model (FRDM) \cite{frdm97,frdm16}, Global Nuclear Mass Model (WS3) \cite{ws3} predictions and the experimental data \cite{wang12} in Table \ref{tab1} and also display in Fig. \ref{fig1} for binding energy per nucleon. We have not given to keep the clarity of the figure.\\

Table \ref{tab1}, one can notice that the calculated values obtained from RMF for NL3 $\&$ NL3$^*$, and DD-ME2 parameters seem to be in reasonably good agreement with the FRDM $\&$ WS3 predictions and available experimental data. We have also estimated the standard deviation of the results obtained from RMF for NL3, NL3$^*$, DD-ME2 parameter sets, FRDM, and WS3 predictions with-respect-to (w.r.t) the available experimental data. Quantitatively, the standard deviation of NL3, NL3$^*$, and DD-ME2 parameter sets are 1.5342, 1.9978, and 1.6282, respectively.  In the case of FRDM and WS3, the magnitude of the standard deviation is a little bit lower value (1.2064 and 1.1982) compared to the RMF results. This higher deviation is obvious for self-consistent microscopic models \cite{ring90,lala99c,lala09}. From Fig. \ref{fig1}, we notice that in the isotopic chains of $Z$ = 120 to 112, the binding energy per particle (BE/A) increases slowly with the decrease of mass number whereas, for the isotopic chains of $Z$ = 110 to 100, there is a rise with the decrease in mass number. Although the standard deviation obtained from NL3 parameter sets for these considered nuclei is slightly lower than the NL3$^*$ and DD-ME2 parameter sets, those are the refitted and density-dependent versions of the NL3 parameter set by adopting various updated experimental data from finite nuclei and infinite nuclear matter. For example, the NL3$^*$ and DD-ME2 parameter set is refitted by considering the new experimental binding energy data, more reliable information about the neutron skin, correct nuclear sizes, and updated infinite nuclear matter constraints. More details of the parameter and its importance can be found in Ref. \cite{niks05,niks11,lala09}. Hence it is important and also interesting to examine the applicability of the NL3$^*$ and DD-ME2 parameter sets to predict the structural and decay properties of finite nuclei.  \\

The pairing energy offers a credible indication of shell/sub-shell closure in terms of single-particle occupancy. Hence, it is interesting to calculate the pairing energy of these superheavy nuclei. In Fig. \ref{fig2}, we have shown the pairing energy of the nucleus $E_{pair}$ (Upper panel) and neutron pairing energy ($E^n_{pair}$) (Lower panel) for the nuclei involve in the $\alpha$-decay chains of $^{296,298,300,302,304}$120 for NL3 (solid line), NL3$^*$ (dotted line), and DD-ME2 (dashed line) force parameter sets. It is worth mentioning that the constant gap BCS for NL3 and NL3$^*$, and Bogoliubov transformation for DD-ME2 parameter sets are adopted to take care of pairing correlations. The results for the $E_{pair}$ are also listed in Table \ref{tab1}. The upper panel notices that the pairing energy decreases over the isotopic chain for NL3 and NL3$^*$ parameter sets except a few neutron numbers. Similar trends also appear for the neutron pairing energy for these two parameter sets. Contrary to the NL3 and NL3$^*$ predictions, we find relative abnormal changes in the pairing over the isotopic chain of all nuclei for the DD-ME2 parameter set. A careful inspection of the magnitude of relative changes shows that a substantial fall appears at $N$ = 164, 172, 176, 182, and 184 for all atomic nuclei for DD-ME2 parameter sets, which is minute in the case of NL3 and NL3$^*$ parameter sets. The relative fall in the magnitude of pairing energy for certain neutron numbers can be correlated with the shell/sub-shell closer structure. \\

Further comparing the results from two different prescriptions of pairing correlation adopted in the present analysis, we find a clear picture of the trend in pairing energy from Bogoliubov transformation over the BCS approach. It is to be noted that $N$ = 172, 182, and/or 184 are predicted to be the next magic neutron in the superheavy valley \cite{bhu12,zhang05} and reference therein. The present analysis further supports the recent predictions from various phenomenological semi-classical, microscopic non-relativistic, and relativistic models \cite{bhu12,rutz97,bend99,krup00,rein02,bend01,meng96,meng98,zhang05}. It is also found that $E_{pair}$ follows nearly a similar trend for all the isotopic chains except for $Z$ = 106 and $Z$ = 112, where a marginal difference is observed. This difference is due to the proton pairing energy because the trend disappears in neutron pairing. And hence more systematic studies are highly welcome to determine the strength of pairing and its model dependency by adopting various pairing correlations. \\

\noindent
\subsection{Quadrupole Deformation and Nuclear Radius}
The quadrupole deformation parameter $\beta_{2}$ is a measure of the most straightforward deviation from spherical symmetry in the nuclear density distribution, i.e., the shape of the nucleus in their ground and intrinsic excited state. The ground state quadrupole deformation parameter $\beta_{2}$ for RMF formalism using NL3, NL3$^*$, and DD-ME2 force parameter set is displayed in Fig. \ref{fig3} and also listed in Table \ref{tab1}. Further, the ground state root-mean-square charge radius ($r_{ch}$) for all these nuclei are also listed in Table \ref{tab1} for NL3, NL3$^*$, and DD-ME2 parameter sets. We find a similar trend also consistent in magnitude for all three parameter sets. The calculated quadrupole deformation is also compared with the values from FRDM predictions \cite{frdm95}. It is clear from the figure that $\beta_{2}$  values for NL3, NL3$^*$, and DD-ME2 values are in very close agreement with each other and also matching to the FRDM predictions for the isotopes of $Z$ = 100 to 110. For example, the calculated $\beta_2$-value keeps on decreasing as the mass number is increased over the isotopes of $Z$ = 100 to 110 with a range, 0.1 $\le \beta_2 \le$ 0.3, which are consistent with the FRDM predictions. In the case of $Z$ = 112 to 120, the $\beta_2$-value follows a significant jump from light deform prolate configuration to highly/superdeformed prolate shape for both NL3 and NL3$^*$ parameter set. In DD-ME2, the shape transition from prolate to the superdeformed prolate configuration is observed for the isotopes of $Z$ = 118 and 120. In contrast, the FRDM has no shape change at all, i.e., it just follows the same pattern and a few isotopes of slight oblate configuration. Although there is no experimental data available, it will still be interesting to see the superdeformed and/or hyperdeformed ground state followed by a low lying intrinsic excited state of the spherical configuration. Further, this shape transition in the superheavy valley could consider as the traditional trend as seen in some of our previous works \cite{bhu09,bhu12a} and references therein. One can further explore the ground state structure by considering more possible degrees of freedom, namely, octupole and hexadecapole deformations \cite{lu16} and reference therein. 

\begin{figure*}
\begin{center}
\includegraphics[width=0.45\columnwidth]{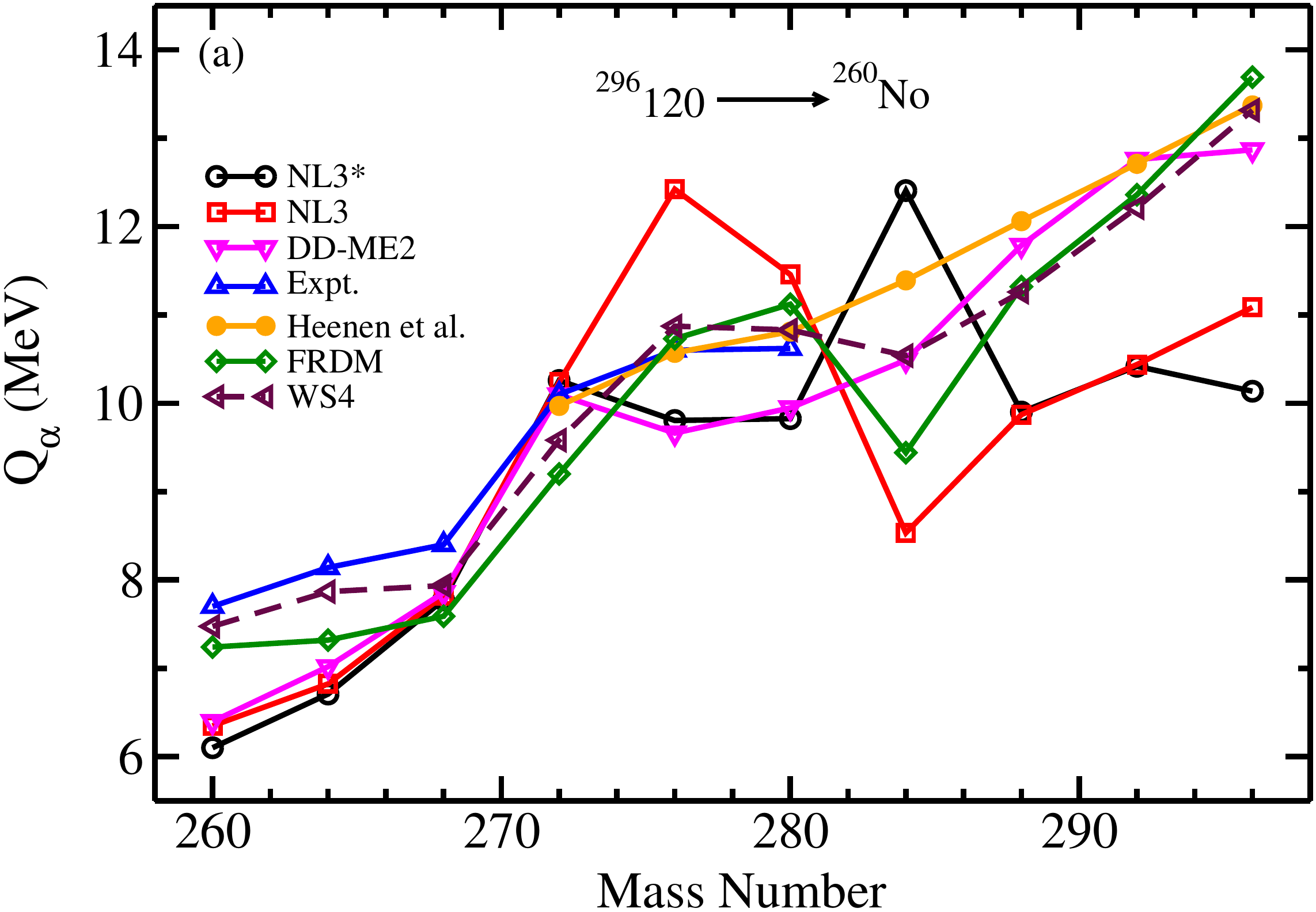}
\includegraphics[width=0.45\columnwidth]{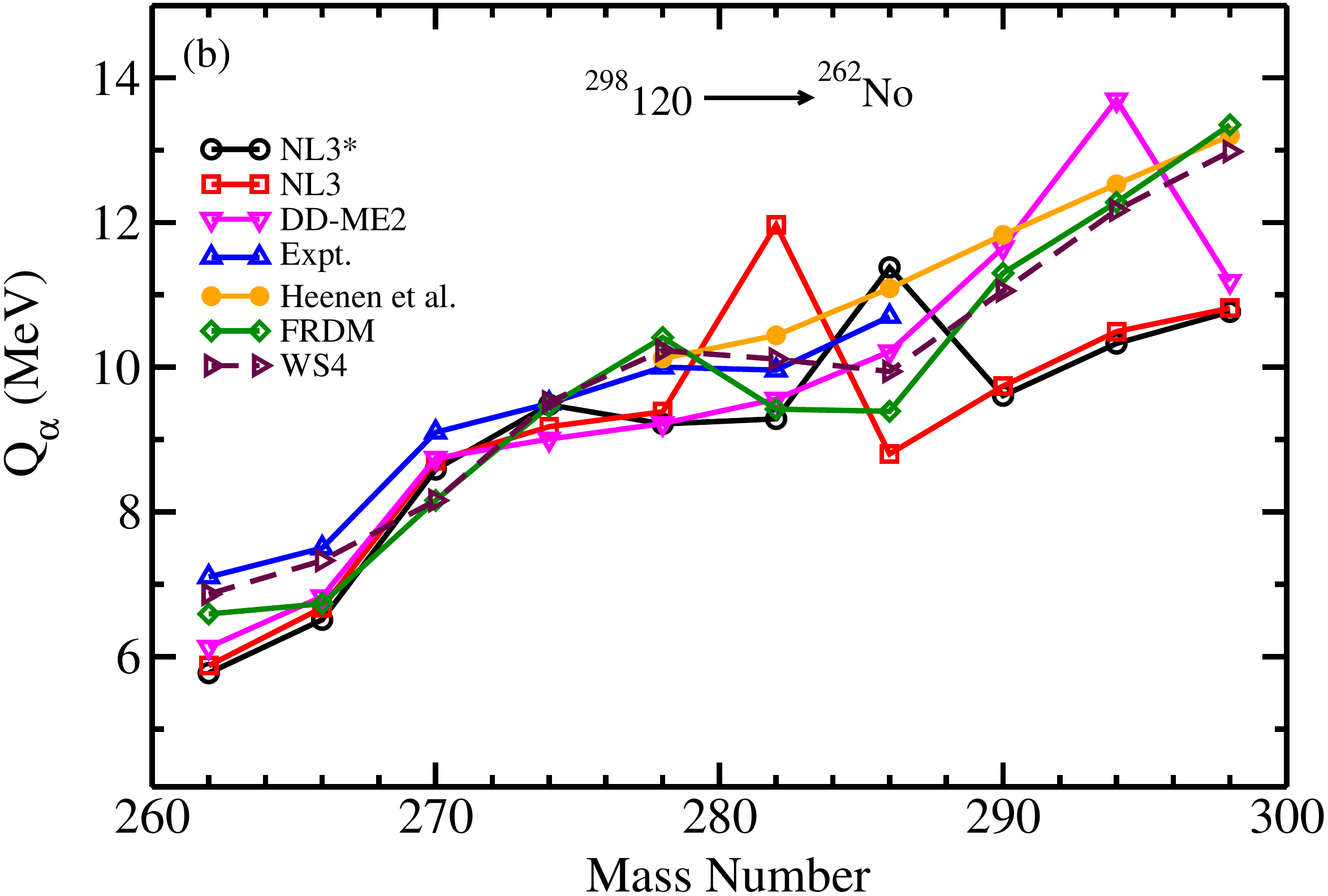}
\includegraphics[width=0.45\columnwidth]{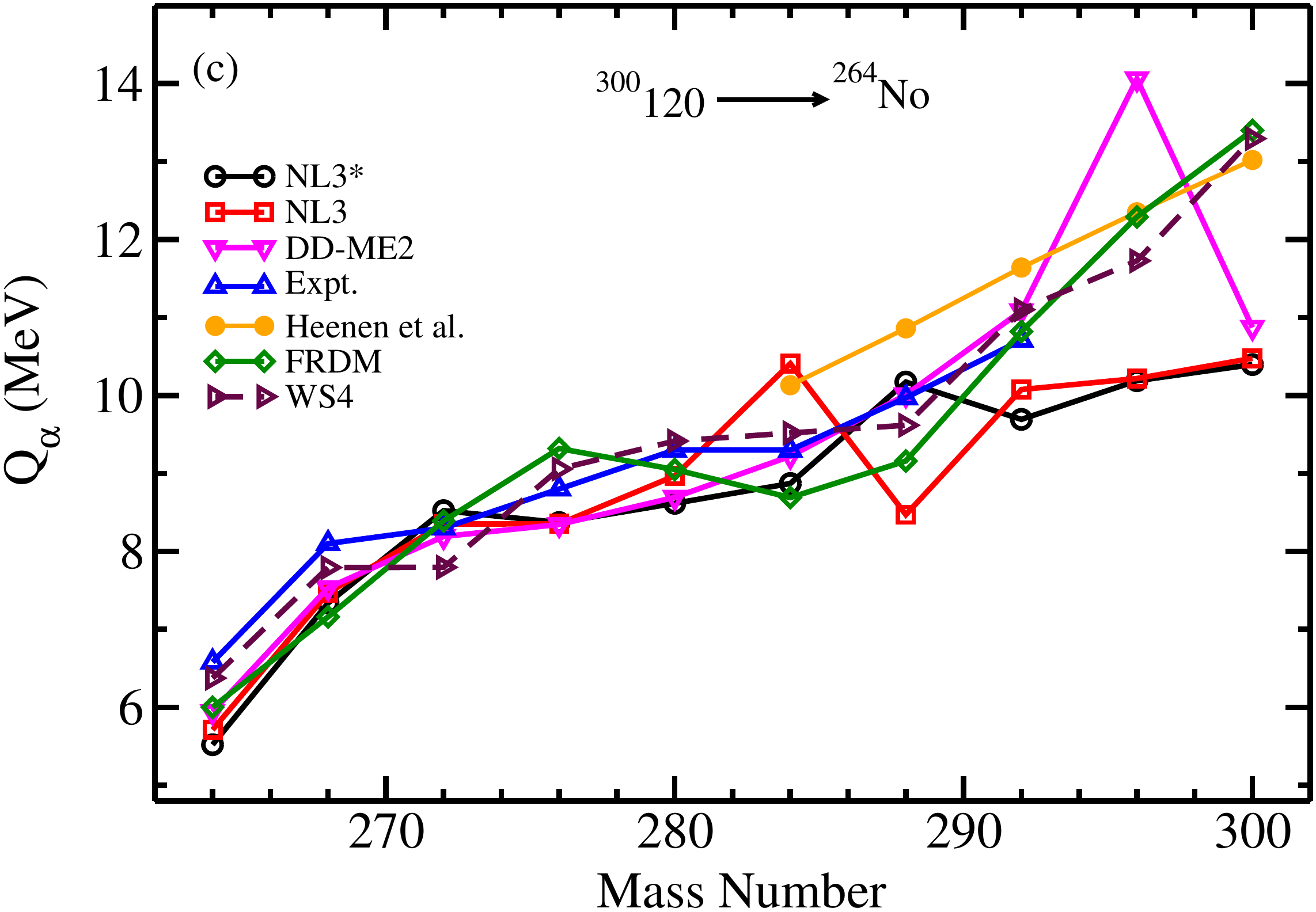}
\includegraphics[width=0.45\columnwidth]{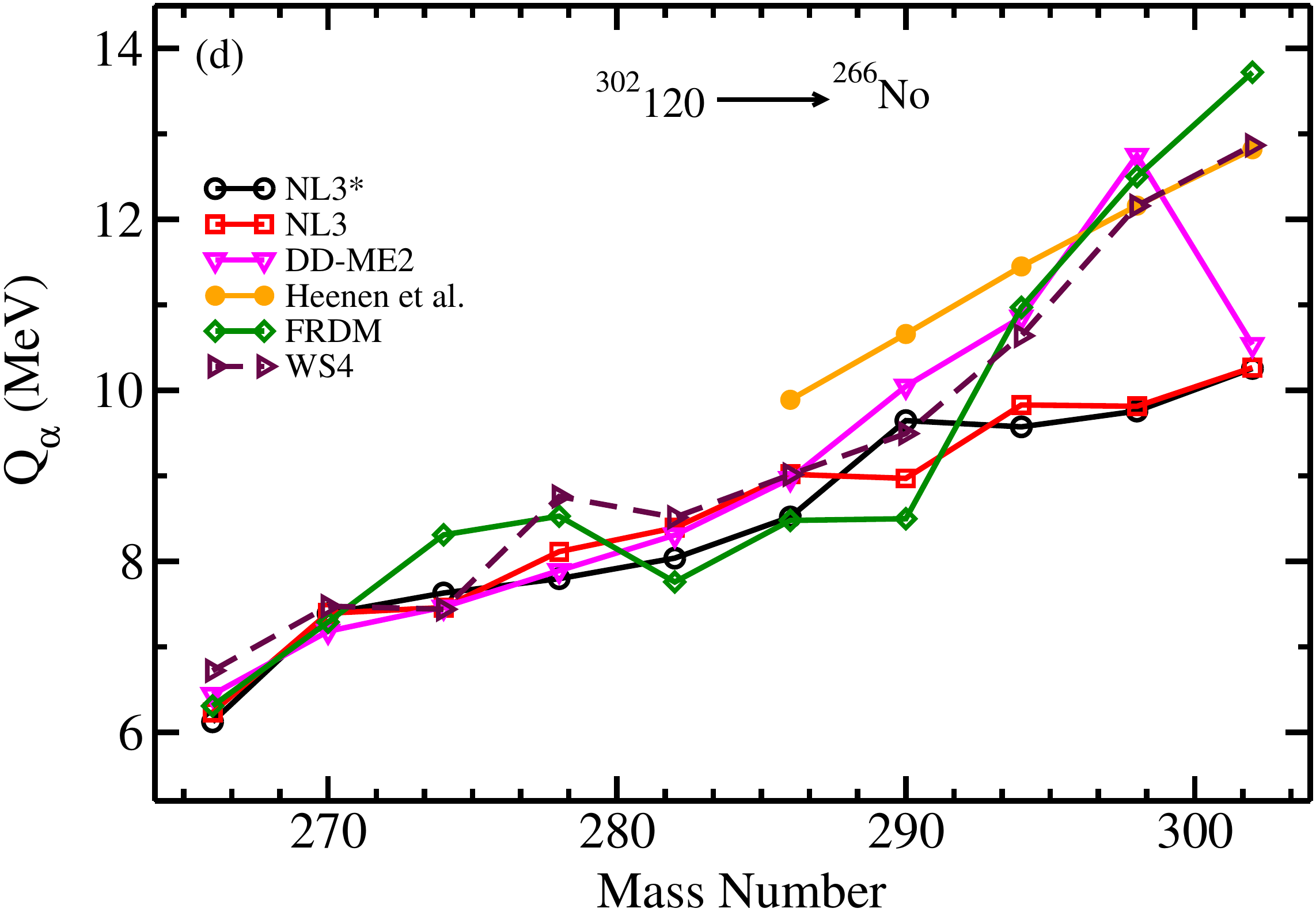}
\includegraphics[width=0.45\columnwidth]{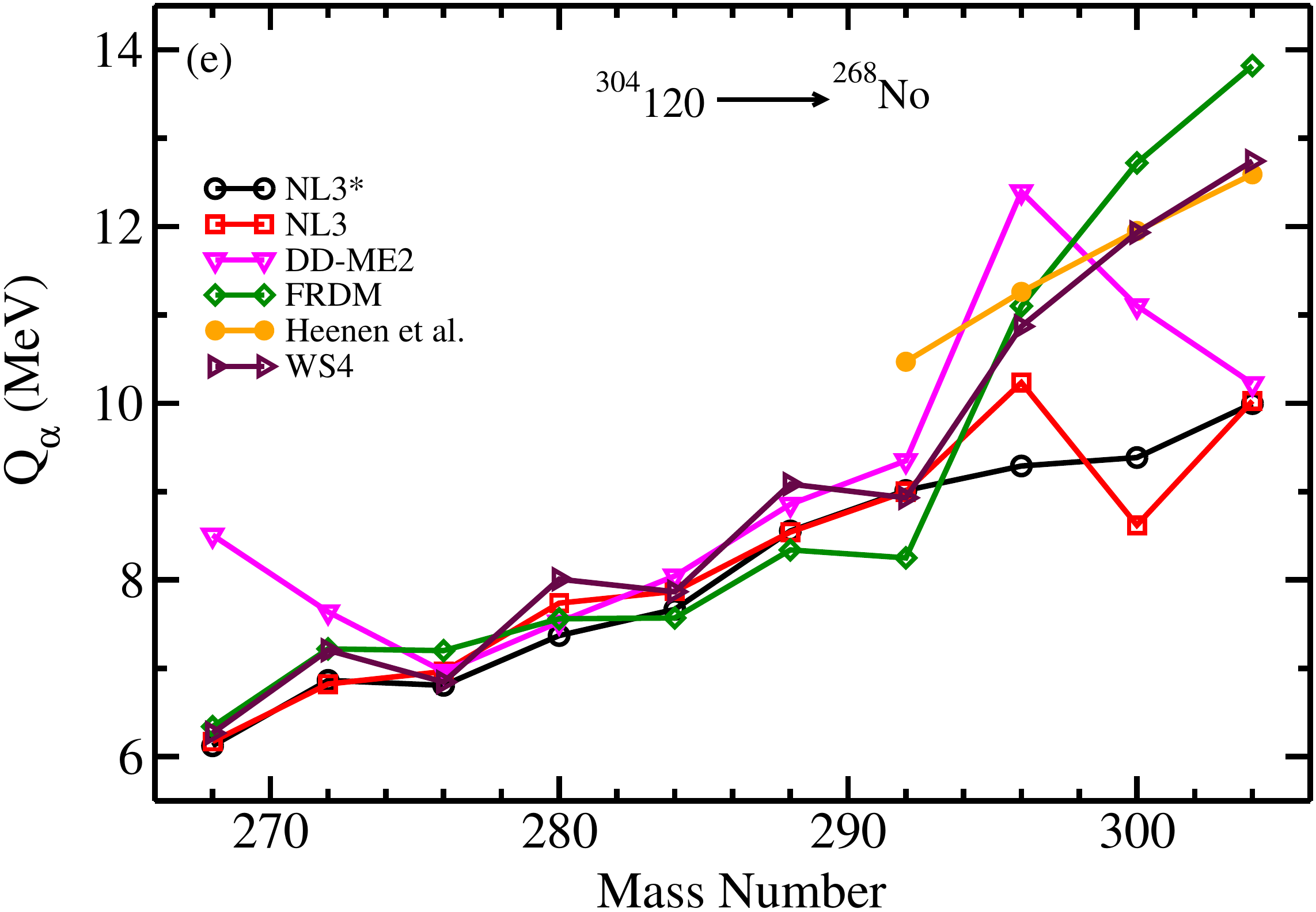}
\caption{\label{fig4}(Color online) The $Q_\alpha$ energy for the $\alpha$-decay chains of (a) $^{296}$120 $\rightarrow$ $^{260}$No, (b) $^{298}$120 $\rightarrow$ $^{262}$No, (c) $^{300}$120 $\rightarrow$ $^{264}$No, (d) $^{302}$120 $\rightarrow$ $^{266}$No, and (e) $^{304}$120 $\rightarrow$ $^{268}$No, calculated using RMF with NL3, NL3$^*$, and DD-ME2 parameter sets are compared with the FRDM \cite{frdm97,frdm16}, Heenen {\it et al.,} \cite{heen15}, WS4 predictions \cite{ws4} and available experimental data \cite{wang12}.}
\end{center}
\end{figure*}
\begin{figure*}
\begin{center}
\includegraphics[width=0.45\columnwidth]{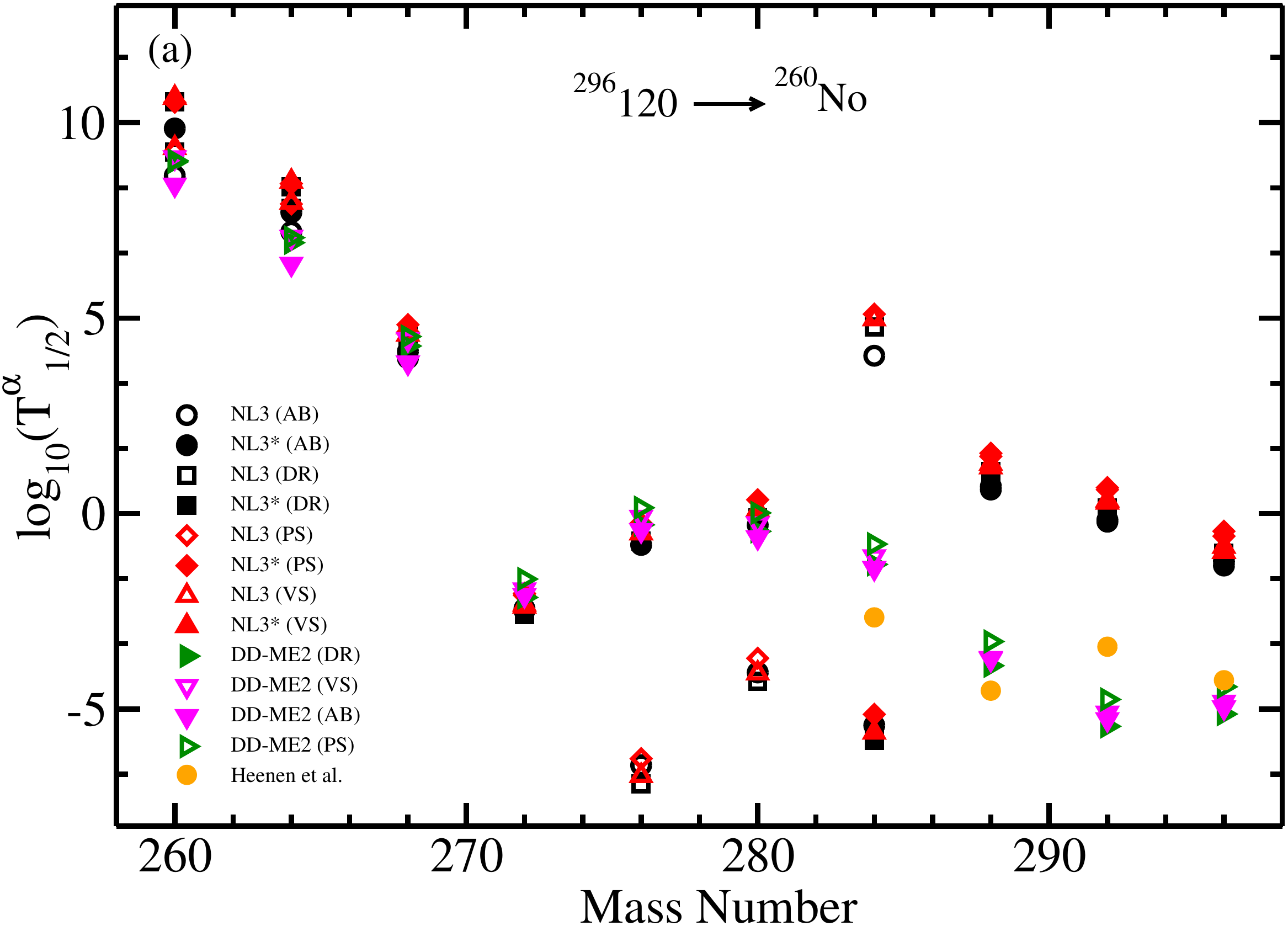}
\includegraphics[width=0.45\columnwidth]{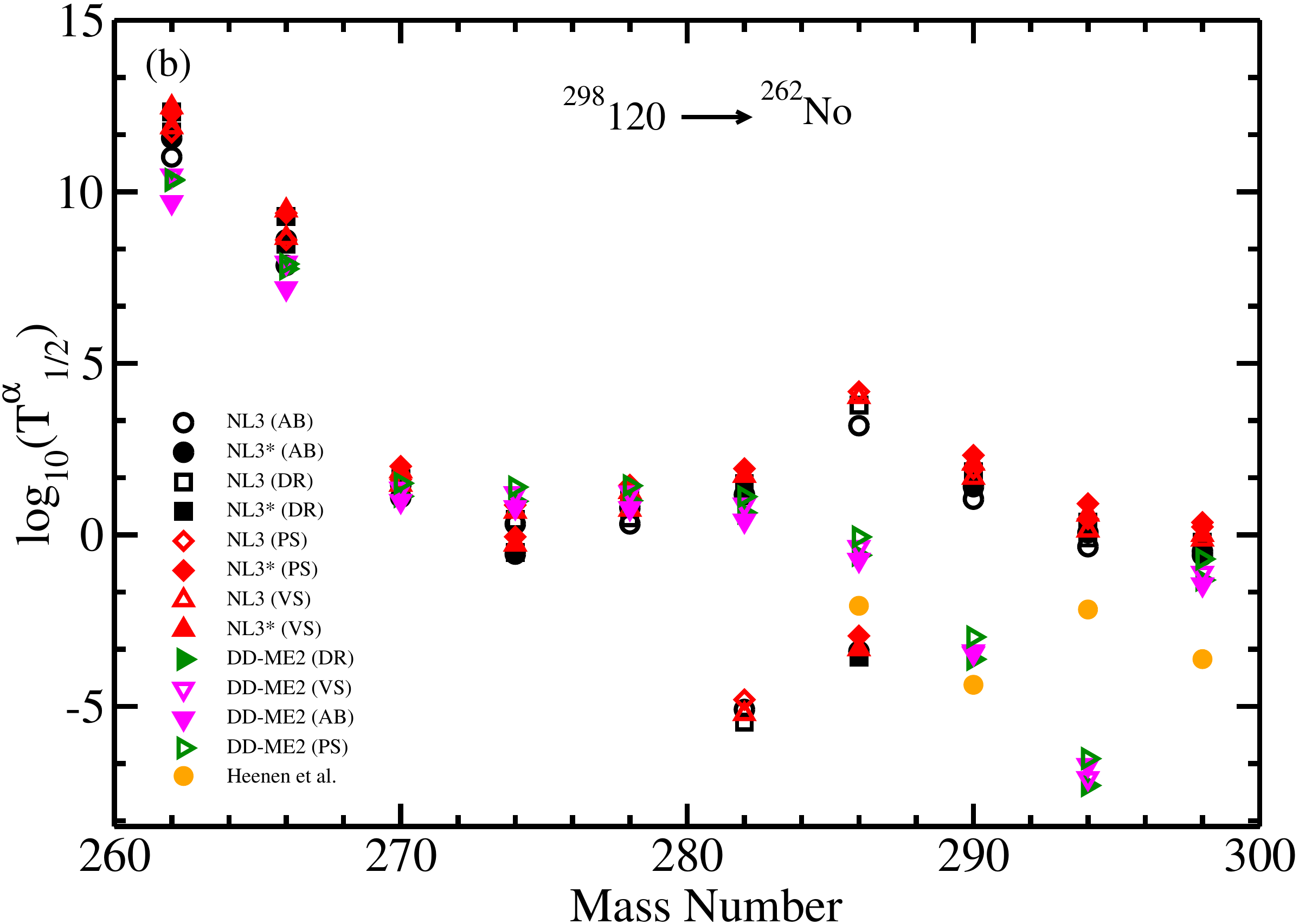}
\includegraphics[width=0.45\columnwidth]{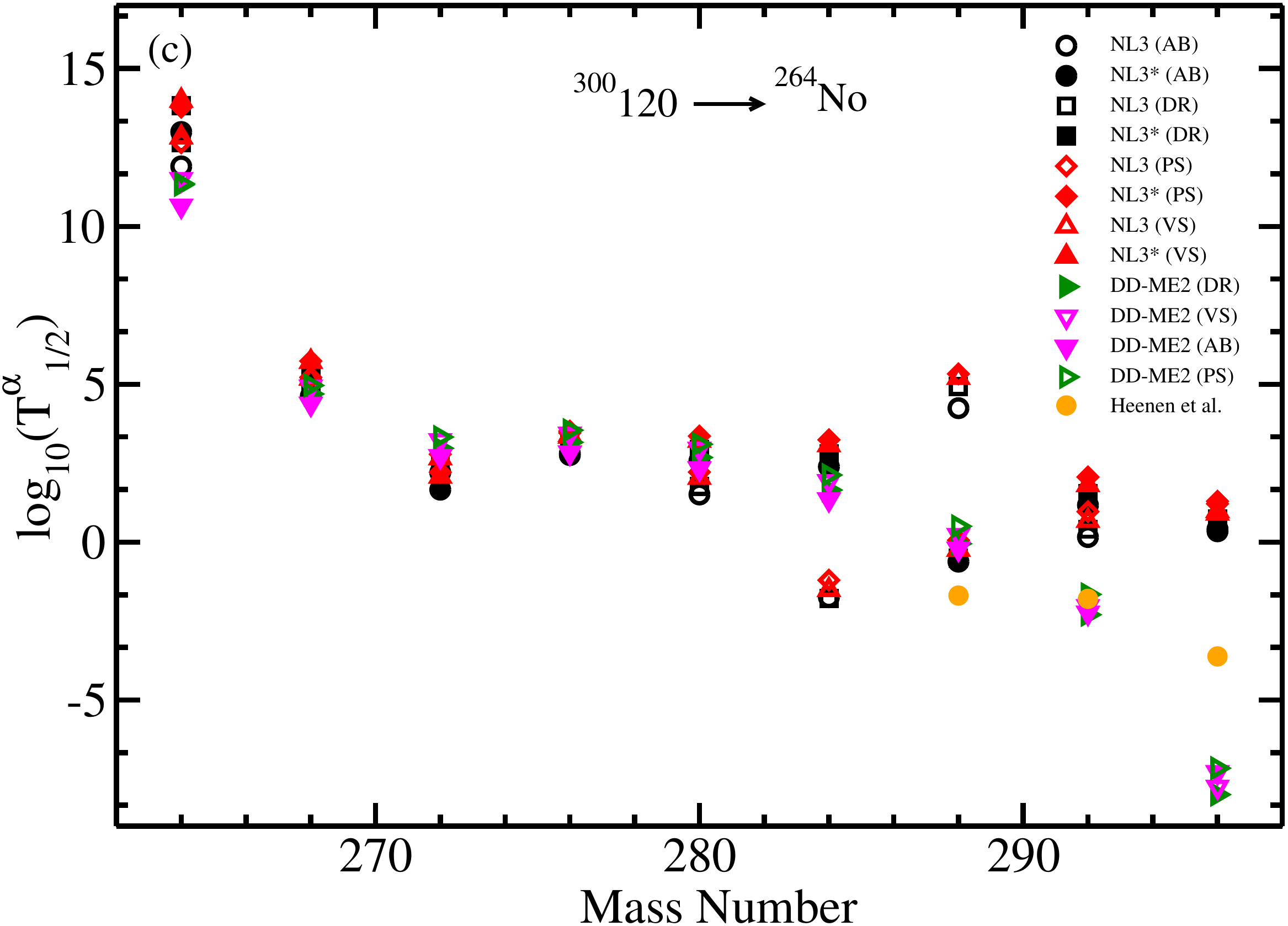}
\includegraphics[width=0.45\columnwidth]{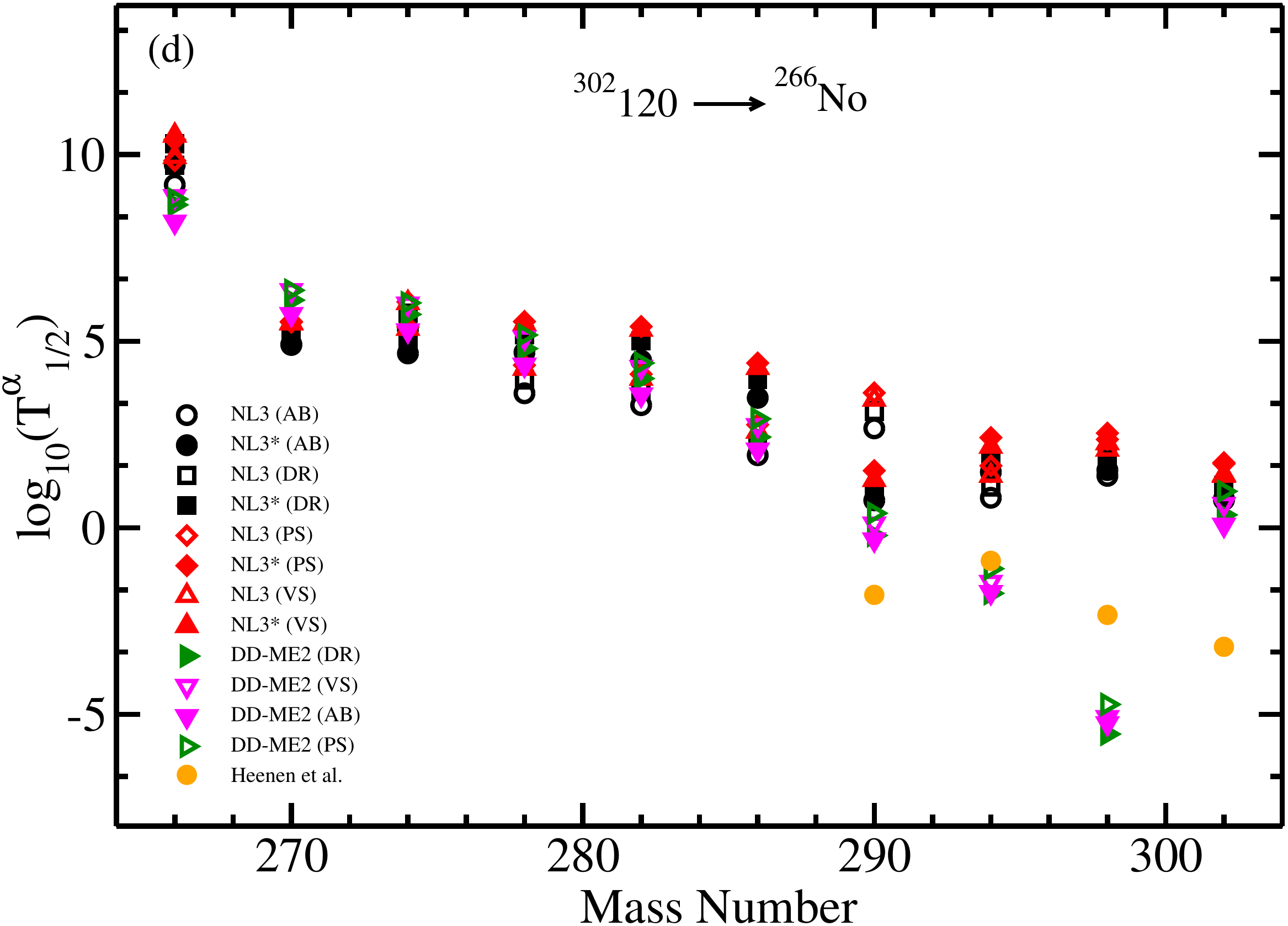}
\includegraphics[width=0.45\columnwidth]{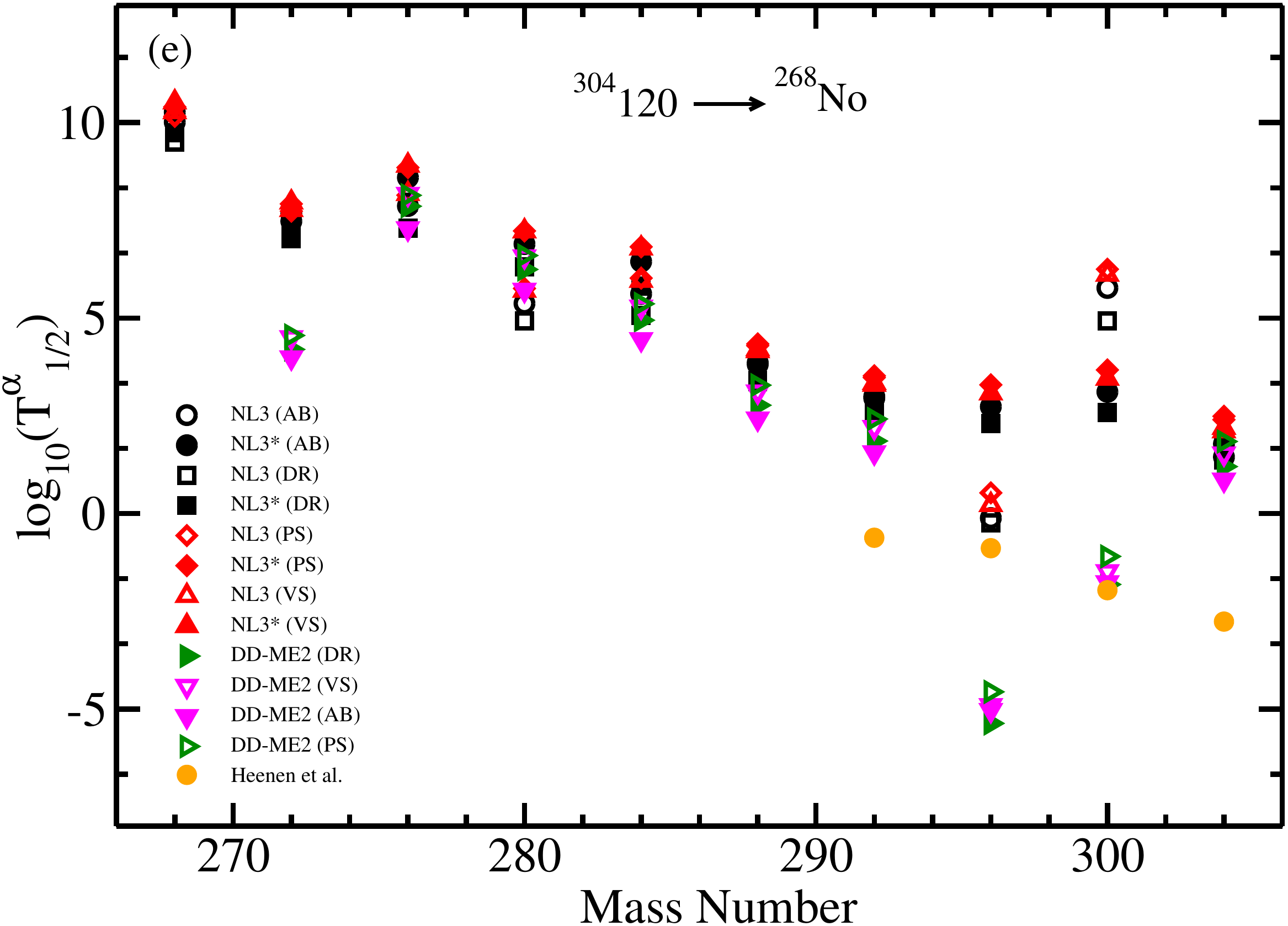}
\caption{\label{fig5} The decay half-life $T_{1/2}^{\alpha}$ (sec) for five different $\alpha$-decay chains, namely, (a) $^{296}$120 $\rightarrow$ $^{260}$No, (b) $^{298}$120 $\rightarrow$ $^{262}$No, (c) $^{300}$120 $\rightarrow$ $^{264}$No, (d) $^{302}$120 $\rightarrow$ $^{266}$No, and (e) $^{304}$120 $\rightarrow$ $^{268}$No, within relativistic mean field approach for NL3, NL3$^*$, and DD-ME2 parameter sets are calculated using formulae of Viola-Seaborg \cite{viola66,sobic1989} (VS), Brown \cite{brown1992} (AB), Parkhomenko-Sobiczewski \cite{menko} (PS), and Royer \cite{royer2000,dasgupta} (DR). The predictions from Heenen {\it et al.,} \cite{heen15} are given for comparisons.}
\end{center}
\end{figure*}

\section{Nuclear Decay Properties}
\label{decay}
The $\alpha$-decay energy $Q_\alpha$ is the fundamental parameter for understanding decay properties and for calculating the half-lives $T_\alpha$ of a nucleus. Further, the nuclei with $\alpha$-decay half-life relatively larger than the neighboring nuclei indicate the stability and/or shell closure. The $Q_\alpha$ is estimated by knowing the binding energies (BE) of the parent and daughter nuclei and that of $^4$He nucleus and is given by, $Q_\alpha (N, Z)$ = BE($N$-2, $Z$-2) + BE(2,2) - BE($N$, $Z$). Here, BE($N$, $Z$), BE($N$-2, $Z$-2), and BE(2, 2) are the binding energies of the parent, daughter, and $^4$He (BE = 28.296 MeV), respectively with neutron number $N$ and proton number $Z$. The ground state $Q_\alpha$ values are estimated for five successive decay chains of $Z$= 120, namely,  (a) $^{296}$120 $\rightarrow$ $^{260}$No, (b) $^{298}$120 $\rightarrow$ $^{262}$No, (c) $^{300}$120 $\rightarrow$ $^{264}$No, (d) $^{302}$120 $\rightarrow$ $^{266}$No, and (e) $^{304}$120 $\rightarrow$ $^{268}$No using RMF formalism for NL3 and NL3$^*$ parameter sets. The calculated values are compared with the available FRDM \cite{frdm97,frdm16} predictions, theoretical estimation of Heenen {\it et al.} \cite{heen15}, WS4 predictions \cite{ws4} and experimental data \cite{wang12}, as shown in Fig. \ref{fig4} and also tabulated in the Table \ref{tab2}. For the larger mass numbers, the $Q_\alpha$ values from FRDM \cite{frdm97}, Heenen {\it et al.} \cite{heen15}, and WS4 predictions \cite{ws4} are significantly increased as compared to the RMF estimations. This compelling can be connected with the shape transition in the ground state configuration, which is absent in the case of FRDM and Hennen {\it et al.,} \cite{heen15} for the higher mass region. Furthermore, without any experimental data for the shape information ($\beta_2$) and Q-values of $\alpha$-decay, it is ambitious to reach any conclusion. \\

We have also calculated the standard deviation in the $Q_{\alpha}$ values for NL3, NL3$^*$, and DD-ME2 parameter sets with respect to the available experimental data. The estimated standard deviation for NL3, NL3$^*$, and DD-ME2 are 1.0249, 0.8272, 0.7645, respectively. These values are a little bit overestimated compared to the FRDM (0.5450) and WS4 estimates (0.5854) that exactly reflected in the binding energy. The estimated $Q_{\alpha}$-values from Heenen {\it et al.,} \cite{heen15} are given for only unknown mass nuclei, where no experimental data are available, hence it is not possible to obtain the standard deviation. A careful observation of the standard deviation for these two-parameter sets show that DD-ME2 and NL3$^*$ are slightly superior to NL3, which is precisely reverse in the case of binding energies. This shows the applicability of DD-ME2 and NL3$^*$ over NL3 in predicting the decay properties of superheavy nuclei, which is the primary motivation behind the refitting of NL3, so-called NL3$^*$ \cite{lala09} and density-dependent DD-ME2 \cite{niks05} parameter set. As a few experimental data is available, hence it is also crucial to predict the unknown region of the superheavy island in terms of decay energies and corresponding half-lives.  \\

\begin{figure}
\vspace{-0.4cm}
\begin{center}
\includegraphics[width=0.9\columnwidth]{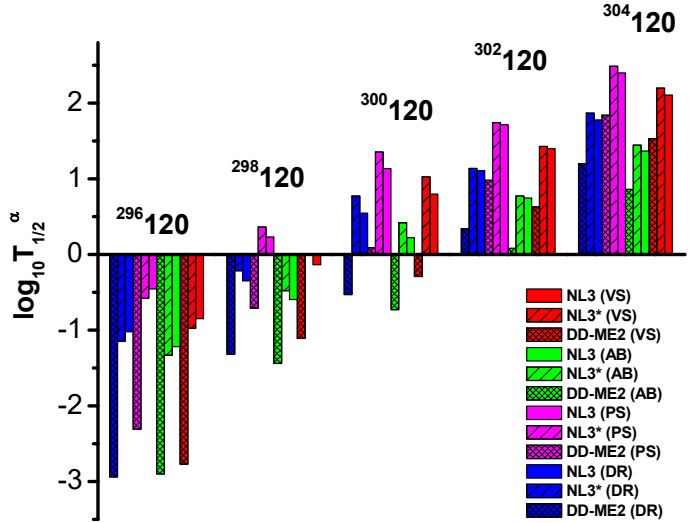}
\vspace{-1.3cm}
\caption{\label{fig6} The decay half-life log$_{10} T_{1/2}^{\alpha}$ (sec) for first successive $\alpha$-decay of five different isotopes of $Z$ =120, namely, $^{296}$120, $^{298}$120, $^{300}$120, $^{302}$120, and $^{304}$120, within the relativistic mean field approach for NL3 (solid bar), NL3$^*$ (stripe bar), and DD-ME2 (dotted bar) parameter sets are calculated using formulae of Viola-Seaborg \cite{viola66,sobic1989} (VS), Brown \cite{brown1992} (AB), Parkhomenko-Sobiczewski \cite{menko} (PS), and Royer \cite{royer2000,dasgupta} (DR), are shown.}
\end{center}
\end{figure}

\noindent
{\bf The $\alpha$-decay half-life ($T_{1/2}^{\alpha}$):}
Four different formulae are employed to estimate the decay half-life. The main motivation behind the adoption of four different $\alpha$-decay formulae is to determine the relative analytic dependency or discrepancy in the decay half-life for a specific Q-value of a particular nucleus. The details are as follows: \\
(i) The expression for the $\alpha$-decay half-life from Viola-Seaborg \cite{viola66} is given by:
\begin{eqnarray}
\log T_{1/2}^\alpha (v) = (aZ+b)Q^{-1/2}_\alpha+(cZ+d)+\textrm{h\textsubscript{log}}. 
\end{eqnarray}
The constants $a$, $b$, $c$ and $d$ are taken from Sobiczewski {\it et al.} \cite{sobic1989}, which are adjustable parameters assigned with the values $a$ = 1.66175, $b$ = -8.5166, $c$ = -0.20228, 
$d$ = -33.9069 and \textrm{h\textsubscript{log}} is the hindrance factor for the nuclei with unpaired nucleons as, 
\begin{eqnarray}
Z= \textrm{even}, N= \textrm{even}, \textrm{h\textsubscript{log}} = 0, \nonumber \\
Z= \textrm{odd}, N= \textrm{even}, \textrm{h\textsubscript{log}} = 0.772, \nonumber \\
Z= \textrm{even}, N= \textrm{odd}, \textrm{h\textsubscript{log}} = 1.066, \nonumber \\
Z= \textrm{odd}, N= \textrm{odd}, \textrm{h\textsubscript{log}} = 1.114.  \nonumber 
\end{eqnarray}
(ii) The expression for $\alpha$-decay from the work of Brown \cite{brown1992} is given by:
\begin{eqnarray}
\log T_{1/2}^\alpha (b) = (9.54Z^{0.6}_d/\sqrt{Q_\alpha})-51.37. 
\end{eqnarray}
Here $Z_d$ is the charge number of the daughter nucleus.\\
(iii) A simple phenomenological formula was proposed by Parkhomenko and Sobiczewski \cite{menko} for calculating the $\alpha$-decay half lives of heavy and superheavy nuclei, and is given as follows:
\begin{eqnarray}
\log T_{1/2}^\alpha (p) = aZ[Q _\alpha (Z, N)-\bar{E_i}]^{-1/2}+bZ+c. 
\end{eqnarray}
The constants $a$, $b$, $c$, are 1.5372, -0.1607 and -36.573, respectively. The parameter $\bar E_i$ is the average excitation energy of the daughter nucleus and varies with odd nucleon number as,
\begin{eqnarray}
Z= \textrm{even}, N= \textrm{even}, \bar E_i = 0, \nonumber \\
Z= \textrm{odd}, N= \textrm{even}, \bar E_i= 0.112, \nonumber \\
Z= \textrm{even}, N= \textrm{odd}, \bar E_i= 0.171, \nonumber \\
Z= \textrm{odd}, N= \textrm{odd}, \bar E_i= 0.283.\nonumber
\end{eqnarray}
The above given parameters are obtained by considering the electron screening effect in the energy.\\
(iv) The semi-empirical formula given by Royer \cite{royer2000,dasgupta} is given by: 
\begin{eqnarray}
\log T_{1/2}^\alpha (r) = a + b A^{1/6}Z^{1/2} + \frac{c Z}{Q_\alpha^{1/2}}.
\end{eqnarray}
The constants $a$, $b$ and $c$ have different values for various combination of charge number $Z$ and mass number $A$ of the parent nuclei, as given below: 
\begin{eqnarray}
Z= \textrm{even}, N= \textrm{even}, \bar a=-25.31, b=-1.1629, c=1.5864, \nonumber \\
Z= \textrm{odd}, N= \textrm{even}, \bar a=-25.68, b=-1.1423, c=1.592, \nonumber \\
Z= \textrm{even}, N= \textrm{odd}, \bar a=-26.65, b=-1.0859, c=1.5848, \nonumber \\
Z= \textrm{odd}, N= \textrm{odd}, \bar a=-29.48, b=-1.113, c=1.6971. \nonumber 
\end{eqnarray}
Royer \cite{royer2000} had described the potential barrier in $\alpha$-decay in terms of quasi-molecular-shape path within the generalized liquid drop model including the proximity effects between nucleons in a neck and the shell effects given by the droplet model. The uncertainties in the calculated half-lives due to the semi-empirical approach are far smaller than the uncertainties due to errors in the calculated energy release. \\

The above mentioned four empirical formulae are known for estimating/ predicting the $\alpha$-decay half-lives in superheavy mass region \cite{viola66,sobic1989,brown1992,menko,royer2000,dasgupta}. The calculated half-lives for the five different $\alpha$-decay chains of different isotopes of $Z$=120, namely, (a) $^{296}$120 $\rightarrow$ $^{260}$No, (b) $^{298}$120 $\rightarrow$ $^{262}$No, (c) $^{300}$120 $\rightarrow$ $^{264}$No, (d) $^{302}$120 $\rightarrow$ $^{266}$No, and (e) $^{304}$120 $\rightarrow$ $^{268}$No, from these formulae are shown in Fig. \ref{fig5} and tabulated in Table \ref{tab2}. It is observed that a little variation is there in the half-lives calculated using the above-said formulae for NL3, NL3$^*$, and DD-ME2 force parameter sets. As mentioned in the previous subsection, the variation is small in $Q_\alpha$, and these values are the input for the analytical formulae of half-life, so the log $T_{1/2}^\alpha$ also appears to be in line with the $Q_\alpha$ variation. Also, it is clear from Table \ref{tab2} that for heavier mass numbers, the log $T_{1/2}^\alpha$ values are significantly different for RMF and FRDM. However, with relatively lower mass numbers for the nuclei, the RMF and FRDM values are both in reasonable agreement. It is worth noting here that for estimating half-lives for the FRDM case \cite{frdm97,frdm16}, the analytical formula of Viola-Seaborg was employed \cite{viola66}. In the recent works of Heenen {\it et al.,} \cite{heen15} and Wang {\it et al.} \cite{wang13}, the $\alpha$-decay half-life study of various isotopes of $Z$ = 120 have been performed. The predicted results from these works \cite{heen15,wang13} reasonably agree with the present analysis. More attentive observation on the half-life of a specific nucleus shows that for a specific $Q_\alpha$ value of a nucleus, there is a substantial difference in the half-life estimates from these four formulae. For example, for $^{304}$120, the $Q_{\alpha}$ value is 10.03 MeV, for the same value, we obtain four different value of log $T_{1/2}^\alpha$ such as 2.11, 1.37, 2.40, and 1.78 by using the formulae of Viola-Seaborg \cite{viola66,sobic1989} (VS), Brown \cite{brown1992} (AB), Parkhomenko-Sobiczewski \cite{menko} (PS), and Royer \cite{royer2000,dasgupta} (DR), respectively. All log $T_{1/2}^\alpha$ values for $^{304}$120 are significantly different from each other by using four formulae. A similar observation can be drawn for all nuclei from different chains. Hence, it concludes a firm dependency of the half-life on the $\alpha$-decay formula in terms of $Q_{\alpha}$-values for all considered decay chains. \\
In the previous analysis, \cite{bhu12} and reference therein, $Z$ = 120 with $N$ = 182 and/or 184 was predicted as a magic isotope (s) in the superheavy valley by observing the structural observable such as pairing gap, paring energy, nucleon separation energy, and shell correction parameter within the non-relativistic Skyrme-Hartree-Fock and relativistic mean-field approaches for various parameter sets. Hence it is one of the prime interests to clarify this prediction in terms of decay half-lives. Hence, we compare the decay half-life for the first successive $\alpha$-decay of five isotopes of $Z$ = 120 with $N$ = 176, 178, 180, 182, and 184. The half-life of $^{296}$120, $^{298}$120, $^{300}$120, $^{302}$120, and $^{304}$120 nuclei are calculated for NL3 (solid bar), NL3$^*$ (stripe bar), and DD-ME2 (dotted bar) using the above said formulae and shown in Fig. \ref{fig6}. It is observed from this figure that the half-life of $^{304}$120 is the maximum among the considering five possible chains of $Z$ = 120 irrespective of the force used within all the four decay half-life formulae. This afresh the previous prediction \cite{bhu12,rana20} in terms of decay energy and half-life that $Z$ =120 with $N$ =184 is the double magic nucleus in the superheavy island beyond $^{208}$Pb. Further, we notice a narrow force parameter dependency in the results for decay energy and half-life.
 
\section{Summary and Conclusions}
\label{summary}
We investigated the structural properties such as binding energy (BE), pairing energy ($E_{pair}$), quadrupole deformation parameter ($\beta_2$), and charge radius ($r_{ch}$) for the even-even isotopic chains of $Z$ = 100 - 120 within axially deformed relativistic mean-field formalism (RMF) by employing the NL3, NL3$^*$, and DD-ME2 parameter sets. The $\alpha$-decay properties such as $\alpha$-decay energies ($Q_\alpha$) and corresponding decay half-lives are also estimates for four different empirical formulas. The estimated results are reasonably good in agreement with the available theoretical predictions and the experimental data. We find the standard deviation with respect to available experimental data \cite{wang12} for NL3, NL3$^*$, and DD-ME2 parameter sets, and compare with the  FRDM \cite{frdm97,frdm95,frdm16} and WS3 \cite{ws3,ws4} predictions. The potential energy landscape shows a shape transition from near prolate to superdeformed/hyperdeformed prolate shape appears for the heavier mass nuclei of $Z$ = 112 to 120 within RMF for NL3, NL3$^*$, and $Z$ = 118 $-$ 120 for DD-ME2 parameter sets, which is absent in the case of FRDM prediction. \\

Further, in reference to the synthesis of different isotopes of $Z$ = 120 element and prediction of corresponding decay chains, the half-life ($T_{1/2}^{\alpha}$) for five different $\alpha$-decay chains, $^{296}$120 $\rightarrow$ $^{260}$No, $^{298}$120 $\rightarrow$ $^{262}$No, $^{300}$120 $\rightarrow$ $^{264}$No, $^{302}$120 $\rightarrow$ $^{266}$No, and $^{304}$120 $\rightarrow$ $^{268}$No are estimated, by employing the calculated decay energy ($Q_\alpha$) from RMF for NL3, NL3$^*$, and DD-ME2 parameter sets, using four different analytical formula of Viola-Seaborg \cite{viola66,sobic1989} (VS), Brown \cite{brown1992} (AB), Parkhomenko-Sobiczewski \cite{menko} (PS), and Royer \cite{royer2000,dasgupta} (DR). We observe a discrepancy in the half-life for a specific Q-value of $\alpha$-decay, which proposes for a refitting or improvement in the formula used for half-life calculation in the superheavy region of the nuclear chart. We also observe a minute dependence of the force parameter used in the present analysis. In addition to these, the present study further confirms the prediction of the double magic nucleus is $Z$ = 120 with $N$ = 184 in the superheavy island \cite{bhu12,rana20} and references therein.  Hence, $Z$ = 120 with $N$ =184 is considered to be a suitable candidate for fusion study \cite{bhu20,rana20}, and also in the synthesis process in the laboratory. 

\section*{Acknowledgments}
This work is supported by FOSTECT Project Code: FOSTECT.2019B.04, FAPESP Project Nos. 2017/05660-0, and Board of Research in Nuclear Sciences (BRNS), Department of Atomic Energy (DAE), Govt. of India, Sanction No. 58/14/12/2019-BRNS and by the CNPq - Brazil.

\bigskip
\bigskip

\appendix
\section{Appendices}
\noindent
\begin{landscape}
\begin{table} 
\tbl{The ground state binding energy (BE), quadrupole deformation ($\beta_2$), root-mean-square charge radius (r$_{ch}$) and pairing energy E$_{pair}$ for relativistic mean field using non-linear NL3 and NL3$^*$ parameter sets. The available experimental data \cite{wang12}, finite range droplet model \cite{frdm97,frdm95,frdm16}, WS3 \cite{ws3} predictions are given for comparison.}
{\begin{tabular}{@{}c|cccccc|ccc|cccc|ccc@{}} \toprule
Nucleus & \multicolumn{6}{c|}{Binding energy (MeV)}  & \multicolumn{3}{c|}{r$_{ch}$ (fm)} & \multicolumn{4}{c|}{Deformation ($\beta_2$)} & \multicolumn{3}{c}{E$_{pair}$} \\
 & NL3 & NL3$^*$ & DD-ME2 &WS3 & FRDM & Expt. & NL3 & NL3$^*$ & DD-ME2 & NL3 & NL3$^*$ & DD-ME2 & FRDM & NL3 & NL3$^*$ & DD-ME2 \\
\colrule
$^{296}$120 & 2093.19 &2090.29 &2087.09&  2083.94& 2085.32&         &6.348 &6.329&6.396& 0.542&0.545&0.250&-0.096& 14.76&14.44 &14.52\\ 
$^{298}$120 & 2107.34 &2104.30 &2099.51&2098.46& 2099.73&         &6.402 &6.397&6.425& 0.551&0.554&0.210&-0.079& 14.16&14.01 &15.88\\ 
$^{300}$120 & 2120.92 &2117.67 &2112.10&2111.92 & 2113.39&         &6.351 &6.346&6.454& 0.561&0.564&0.554&-0.008& 13.57&13.33 &9.66\\ 
$^{302}$120 & 2133.81 &2130.28 &2125.42&2125.1& 2126.05&         &6.368 &6.352&6.483& 0.578&0.585&0.570& 0.000& 12.94&12.65 &8.96\\
$^{304}$120 & 2146.28 &2042.55 &2137.52&2137.63& 2137.99 &         &6.391 &6.281&6.512& 0.580&0.587&0.576&0.000 & 13.12&12.87 &8.61\\
$^{292}$118 & 2075.98 &2073.13 &2070.66&2068.86&  2070.72 &        &6.545 &6.543&6.285& 0.526&0.532&0.132&0.081 & 14.81&14.90 &10.80\\
$^{294}$118 & 2089.86 &2086.77 &2082.40&2083.1& 2084.78 &         &6.561 &6.558&6.343& 0.533&0.536&0.204&-0.087& 14.35&14.48 &14.62\\
$^{296}$118 & 2103.10 &2099.77 &2094.68&2096.94& 2098.49 &        &6.581 &6.577&6.375& 0.543&0.545&0.232&-0.079& 13.92&14.06 &16.42\\
$^{298}$118 & 2115.79 &2112.24 &2107.65&2109.71& 2111.47 &         &6.602 &6.597&6.389& 0.554&0.556&0.233&-0.008& 13.63&13.77 &12.01\\
$^{300}$118 & 2128.01 &2124.22 &2119.45&2122.1& 2123.51 &         &6.632 &6.629&6.543& 0.574&0.579&0.472&0.000 & 13.34&13.44 &14.98\\
$^{288}$116 & 2058.12 &2055.26 &2055.13&2052.61& 2054.78 &         &6.501 &6.501&6.267& 0.517&0.525&0.164&-0.104& 14.68&14.80 &5.27\\
$^{290}$116 & 2072.05 &2068.81 &2067.80&2066.8& 2068.76 &2064.800 &6.510 &6.509&6.268& 0.516&0.520&0.143&0.072 & 14.17&14.33 &6.18\\
$^{292}$116 & 2085.02 &2081.66 &2080.44&2080.23& 2082.48 &2079.916 &6.526 &6.524&6.263& 0.520&0.524&0.082&-0.07 & 13.84&13.93 &13.03\\
$^{294}$116 & 2097.30 &2093.70 &2092.11&2093.56& 2095.68 &         &6.545 &6.542&6.270& 0.530&0.532&0.065&-0.044& 13.61&13.76 &13.23\\
$^{296}$116 & 2108.33 &2105.31 &2102.25&2105.74&  2107.94 &        &6.278 &6.562&6.326& 0.540&0.542&0.163&-0.008& 13.29&13.42 &15.35\\
$^{284}$114 & 2039.69 &2036.86 &2038.62&2035.45&2037.81 &         &6.462 &6.465&6.238& 0.512&0.525&0.168& 0.062& 14.61&14.79 &6.89\\
$^{286}$114 & 2053.50 &2050.12 &2051.16&2049.42& 2051.59 &2048.046 &6.466 &6.467&6.245& 0.502&0.510&0.157&-0.096& 14.02&14.24 &6.98\\
$^{288}$114 & 2066.80 &2063.06 &2063.23&2062.88& 2065.01 &2062.368 &6.476 &6.475&6.249& 0.500&0.505&0.143&0.053 & 13.44&13.67 &5.15\\
$^{290}$114 & 2078.84 &2074.98 &2074.67&2075.74& 2078.35 &        &6.490 &6.489&6.249& 0.504&0.510&0.105&-0.026& 13.27&13.46 &11.73\\
$^{292}$114 & 2090.27 &2086.31 &2086.35&2088.25& 2090.75 &         &6.507 &6.504&6.253& 0.511&0.515&0.066&-0.018& 13.12&13.21 &12.84\\
$^{280}$112 & 2019.93 &2020.97 &2020.81&2017.86& 2018.94 &2016.840 &6.322 &6.212&6.212& 0.483&0.489&0.180&0.08  & 14.82&13.79 &10.19\\
$^{282}$112 & 2034.00 &2033.20 &2033.08&2031.1& 2032.68 &2030.480 &6.432 &6.219&6.219& 0.504&0.507&0.169&0.089 & 14.15&13.54 &11.34\\
$^{284}$112 & 2046.97 &2044.93 &2044.94&2044.09& 2045.88 &2043.948 &6.432 &6.225&6.225& 0.490&0.595&0.156&0.089 & 13.55&13.27 &11.72\\
$^{286}$112 & 2059.51 &2056.33 &2056.42&2056.84&2058.56 &         &6.444 &6.231&6.230& 0.489&0.492&0.140&0.081 & 12.96&12.95 &10.43\\
$^{288}$112 & 2070.97 &2067.02 &2067.40&2068.81& 2070.70 &         &6.456 &6.233&6.232& 0.491&0.495&0.107&-0.061& 12.78&12.73 &13.46\\
$^{276}$110 & 2003.09 &2002.50 &2002.46&2000.53&2001.77 &1999.068 &6.276 &6.186&6.185& 0.346&0.350&0.201&0.212 & 14.11&13.75 &11.86\\
$^{278}$110 & 2017.67 &2014.19 &2014.34&2012.96& 2013.8  &2011.886 &6.198 &6.193&6.193& 0.185&0.188&0.186&0.155 & 13.41&13.59 &13.30\\
$^{280}$110 & 2029.08 &2025.51 &2025.87&2025.32&2026.28 &2024.960 &6.205 &6.199&6.200& 0.172&0.175&0.171&0.108 & 13.24&13.31 &14.74\\
$^{282}$110 & 2040.24 &2036.55 &2037.09&2037.54& 2038.74 &         &6.210 &6.205&6.206& 0.156&0.159&0.155&0.108 & 13.01&13.05 &15.31\\
$^{284}$110 & 2051.22 &2047.28 &2047.96&2049.58& 2050.75 &         &6.213 &6.209&6.211& 0.132&0.140&0.137&0.099 & 12.54&12.60 &14.6\\
$^{272}$108 & 1987.22 &1984.01 &1983.83&1983.27& 1984.2  &1981.248 &6.168 &6.170&6.165& 0.228&0.249&0.234&0.222 & 13.32&13.66 &11.6\\
$^{274}$108 & 1998.76 &1995.11 &1995.27&1995.02& 1995.91 &1993.624 &6.172 &6.168&6.169& 0.206&0.211&0.208&0.212 & 13.18&13.38 &13.66\\
$^{276}$108 & 2009.76 &2005.83 &2006.27&2006.54& 2007.03 &2005.968 &6.179 &6.174&6.176& 0.7190&0.194&0.190&0.164 & 13.06&13.25 &15.92\\
$^{278}$108 & 2020.33 &2016.29 &2017.11&2017.69& 2018.21 &         &6.185 &6.180&6.181& 0.174&0.177&0.172&0.108 & 12.98&13.05 &15.52\\
$^{280}$108 & 2030.79 &2026.65 &2027.71&2029.06& 2030.02 &         &6.189 &6.184&6.187& 0.154&0.159&0.154&0.108 & 12.64&12.68 &15.44\\
$^{268}$106 & 1969.16 &1965.97 &1965.63&1964.58& 1965.11 &1963.100 &6.150 &6.144&6.142& 0.267&0.268&0.268&0.231 & 12.57&12.73 &13.59\\
$^{270}$106 & 1979.64 &1976.29 &1975.98&1976.31& 1977.08 &1975.050 &6.159 &6.154&6.152& 0.258&0.261&0.255&0.221 & 12.92&13.02 &11.16\\
$^{272}$106 & 1989.81 &1985.90 &1986.31&1987.32& 1988.05 &1986.688 &6.154 &6.154&6.153& 0.212&0.227&0.217&0.201 & 12.84&13.12 &14.17\\
$^{274}$106 & 2000.15 &1995.80 &1996.70&1998.15& 1998.44 &        &6.158 &6.155&6.157& 0.189&0.197&0.193&0.164 & 12.66&12.94 &15.54\\
$^{276}$106 & 2010.23 &2005.73 &2006.94&2008.65& 2009.29 &         &6.163 &6.159&6.161& 0.171&0.175&0.170&0.135 & 12.40&12.59 &15.97\\
$^{264}$104 & 1948.69 &1945.45 &1945.19&1944.32& 1944.40 &1943.304 &6.118 &6.112&6.113& 0.274&0.276&0.277&0.22  & 12.53&12.67 &8.74\\
$^{266}$104 & 1960.06 &1956.58 &1956.42&1956.62& 1956.95 &1955.632 &6.127 &6.121&6.121& 0.268&0.269&0.267&0.23  & 12.37&12.48 &7.03\\
$^{268}$104 & 1969.87 &1966.13 &1966.21&1967.54& 1968.14 &1966.584 &6.136 &6.132&6.132& 0.258&0.262&0.255&0.221 & 12.63&12.74 &13.93\\
$^{270}$104 & 1979.31 &1975.13 &1975.87&1977.94& 1978.45 &         &6.137 &6.135&6.135& 0.224&0.237&0.224&0.201 & 12.54&12.72 &14.18\\
$^{272}$104 & 1988.90 &1984.24 &1985.61&1987.97& 1988.19 &         &6.137 &6.136&6.138& 0.190&0.204&0.195&0.164 & 12.24&12.45 &15.79\\
$^{260}$102 & 1927.22 &1923.86 &1923.91&1923.7& 1923.43 &1923.220 &6.083 &6.078&6.08& 0.276&0.279&0.280&0.228 & 12.62&12.75 &11.77\\
$^{262}$102 & 1938.44 &1934.80 &1934.96&1935.65& 1935.38 &1934.870 &6.093 &6.088&6.090& 0.271&0.274&0.274&0.219 & 12.33&12.50 &12.31\\
$^{264}$102 & 1949.04 &1945.17 &1945.45&1947.04& 1947.01 &1946.472 &6.102 &6.097&6.099& 0.264&0.266&0.265&0.22  & 12.05&12.19 &10.41\\
$^{266}$102 & 1958.41 &1954.22 &1954.76&1957.19& 1957.45 &         &6.110 &6.106&6.108& 0.250&0.256&0.249&0.22  & 12.14&12.26 &14.64\\
$^{268}$102 & 1967.43 &1962.81 &1964.95&1966.92& 1967.11 &        &6.114 &6.111&6.113& 0.223&0.230&0.222&0.201 & 11.96&12.09 &15.98\\
$^{256}$100 & 1905.27 &1901.67 &1902.01&1902.75& 1902.38 &1902.538 &6.048 &6.043&6.047& 0.275&0.279&0.281&0.227 & 12.66&12.85 &14.58\\
$^{258}$100 & 1916.01 &1912.27 &1912.79&1914.11& 1913.67 &1913.844 &6.058 &6.053&6.058& 0.273&0.276&0.278&0.228 & 12.40&12.54 &14.88\\
$^{260}$100 & 1926.45 &1922.39 &1923.09&1925.15& 1924.71 &1924.780 &6.068 &6.063&6.068& 0.267&0.269&0.270&0.219 & 12.03&12.19 &15.64\\
$^{262}$100 & 1936.35 &1932.05 &1932.91&1935.66& 1935.47 &        &6.077 &6.072&6.078& 0.258&0.261&0.261&0.22  & 11.71&11.84 &14.16\\
$^{264}$100 & 1945.30 &1940.63 &1945.16&1944.89& 1945.16 &         &6.084 &6.080&6.087& 0.239&0.246&0.230&0.22  & 11.61&11.79 &13.01\\
\botrule
\hline
\end{tabular}
\label{tab1}}
\end{table}
\end{landscape}
\begin{landscape}
\begin{table}[h]
\tbl{The Q$_\alpha$ in MeV for the decay chains of (a) $^{296}$120 $\rightarrow$ $^{260}$No, (b) $^{298}$120 $\rightarrow$ $^{262}$No, (c) $^{300}$120 $\rightarrow$ $^{264}$No, (d) $^{302}$120 $\rightarrow$ $^{266}$No, and (e) $^{304}$120 $\rightarrow$ $^{268}$No for NL3, NL3$^*$, and DD-ME2 parameter sets. The corresponding $\alpha$-decay half lives (sec) in terms of log T$_{1/2}^{\alpha}$ calculated using four different empirical relations from Viola-Seaborg (VS) \cite{viola66,sobic1989}, Brown (AB) \cite{brown1992}, Parkhomenko-Sobiczewski (PS) \cite{menko}, and Royer-Dasgupta-Reyes (DR) \cite{royer2000,dasgupta}, are listed. The available FRDM predictions \cite{frdm97,frdm16}, $Q_{\alpha}$ values from WS4 \cite{ws4} prediction and experimental data \cite{wang12} are given for comparison.}
{\begin{tabular}{@{}|c|cccccc|cccc|cccc|cccc|c|@{}} \toprule
Nucleus & \multicolumn{6}{c|}{Q$_\alpha$ (MeV)} & \multicolumn{13}{c|}{logT$_{1/2}^{\alpha}$} \\
&&&&&&& \multicolumn{4}{|c|}{RMF (NL3)} 
& \multicolumn{4}{c|}{RMF (NL3$^*$)} 
& \multicolumn{4}{c|}{RMF (DD-ME2)} 
& \multicolumn{1}{c|}{FRDM} \\
 & (NL3)& (NL3$^*$) & DD-ME2 &(WS4)& FRDM & Expt. & (VS) & (AB) & (PS) & (DR) & (VS) & (AB) & (PS) & (DR) & (VS) & (AB) & (PS) & (DR) & \cite{frdm97,frdm16}\\
\colrule
$^{296}$120&11.09&11.14	&11.87&13.32& 13.69 &		&-0.85	&-1.22	&-0.46	&-1.02	&-0.98	&-1.33	&-0.58	&-1.15	&-2.77&-2.90&-2.31&-2.94& -6.59 \\
$^{298}$120&10.82&10.77	&11.19&12.98& 13.35 &		&-0.14	&-0.60	&-0.23	&-0.35	&0.00	&-0.48	&0.36	&-0.21	&-1.11&-1.44&-0.71&-1.32&	-5.93 \\
$^{300}$120&10.48&10.40	&10.87&13.29& 13.4	 &		&0.80	&0.22	&1.13	&0.55	&1.02	&0.42	&1.35	&0.77	&-0.29&-0.73&0.09&-0.53&   -6.03 \\
$^{302}$120&10.27&10.26	&10.54&12.87& 13.72 &		&1.40	&0.75	&1.71	&1.11	&1.43	&0.77	&1.74	&1.14	&0.63&0.08&0.98&0.34&	-6.63 \\
$^{304}$120&10.03&10.00	&10.22&12.74& 13.82 &		&2.11	&1.37	&2.40	&1.78	&2.19	&1.44	&2.48	&1.86	&1.53&0.86&1.84&1.20&	-6.83 \\
$^{292}$118&10.44&10.42	&12.76&12.21& 12.36 &		&0.28	&-0.21	&0.61	&0.10	&0.34	&-0.16	&0.66	&0.15	&-5.27&-5.10&-4.76&-5.44&	-4.43 \\
$^{294}$118&10.49&10.33	&13.70&12.17& 12.28 &		&0.13	&-0.35	&0.46	&-0.10	&0.59	&0.06	&0.91	&0.37	&-7.09&-6.71&-6.52&-7.30&	-4.24 \\
$^{296}$118&10.22&10.19	&14.06&11.73& 12.29 &		&0.91	&0.34	&1.22	&0.65	&0.99	&0.41	&1.29	&0.73	&-7.75&-7.29&-7.16&-8.00&	-4.27 \\
$^{298}$118&9.81 &9.76	&12.75&12.16& 12.5	 &		&2.10	&1.39	&2.37	&1.80	&2.27	&1.54	&2.53	&1.97	&-5.24&-5.08&-4.73&-5.53&	-4.73 \\
$^{300}$118&8.62 &9.39	&11.10&11.93& 12.72 &		&6.11	&4.93	&6.25	&5.77	&3.45	&2.58	&3.67	&3.11	&-1.48&-1.77&-1.10&-1.81&	-5.18 \\
$^{288}$116&9.87 &9.90	&11.79&11.26& 11.32 &		&1.27	&0.69	&1.54	&1.08	&1.19	&0.61	&1.46	&0.99	&-3.70&-3.73&-3.27&-3.89&	-2.61 \\
$^{290}$116&9.74 &9.61	&11.65&11.06& 10.12 &11.3	&1.67	&1.04	&1.92	&1.43	&2.08	&1.40	&2.32	&1.84	&-3.40&-3.45&-2.98&-3.63&	-2.12 \\
$^{292}$116&10.08&9.69	&11.09&11.1& 10.82 &10.71	&0.68	&0.16	&0.96	&0.41	&1.81	&1.17	&2.06	&1.54	&-2.03&-2.24&-1.66&-2.30&	-1.73 \\
$^{294}$116&9.83 &9.57	&10.86&10.64& 10.97 &		&1.40	&0.80	&1.66	&1.09	&2.17	&1.49	&2.41	&1.87	&-1.45&-1.73&-1.10&-1.76&	-1.75 \\
$^{296}$116&10.23&9.29	&12.40&10.87& 11.1	 &		&0.23	&-0.24	&0.53	&-0.12	&3.08	&2.30	&3.29	&2.74	&-5.03&-4.91&-4.56&-5.37&	-2.08 \\
$^{284}$114&8.53 &12.41	&10.49&10.54& 9.44	 &10.7	&4.97	&4.04	&5.10	&4.77	&-5.60	&-5.42	&-5.14	&-5.80	&-1.10&-1.40&-0.78&-1.30&	1.93 \\
$^{286}$114&8.80 &11.38	&10.21&9.94& 9.39	 &10.7	&4.02	&3.18	&4.18	&3.78	&-3.33	&-3.39	&-2.94	&-3.57	&-0.36&-0.73&-0.06&-0.59&	2.08 \\
$^{288}$114&8.47 &10.17	&10.01&9.62& 9.16	 &9.97	&5.20	&4.24	&5.32	&4.93	&-0.23	&-0.62	&0.06	&-0.51	&0.22&-0.21&0.50&-0.05&	2.8 \\
$^{290}$114&8.97 &9.65	&10.05&9.5& 8.5	 &		&3.44	&2.66	&3.62	&3.13	&1.28	&0.74	&1.53	&0.97	&0.11&-0.32&0.39&-0.20&	5.08 \\
$^{292}$114&9.00 &9.01	&9.35&8.93& 8.25	 &		&3.34	&2.58	&3.52	&3.00	&3.30	&2.54	&3.48	&2.95	&2.20&1.56&2.42&1.86&	6.02 \\
$^{280}$112&11.46&9.83	&9.95&10.83& 11.12 &10.62	&-4.09	&-4.07	&-3.70	&-4.29	&0.09	&-0.30	&0.35	&-0.11	&-0.25&-0.61&0.01&-0.46&	-3.3 \\
$^{282}$112&11.97&9.29	&9.56&10.11& 9.42	 &9.96	&-5.22	&-5.09	&-4.80	&-5.46	&1.72	&1.17	&1.93	&1.48	&0.88&0.41&1.11&0.64&	1.3 \\
$^{284}$112&10.41&8.87	&9.22&9.52& 8.69	 &9.3	&-1.51	&-1.75	&-1.21	&-1.79	&3.07	&2.39	&3.24	&2.80	&1.92&1.35&2.13&1.65&	3.67 \\
$^{286}$112&9.02 &8.52	&8.97&9.01& 8.48	 &		&2.58	&1.94	&2.76	&2.27	&4.29	&3.48	&4.42	&3.97	&2.74&2.09&2.92&2.43&	4.44 \\
$^{288}$112&8.54 &8.55	&8.86&9.09& 8.34	 &  	&4.21	&3.41	&4.34	&3.86	&4.16	&3.37	&4.29	&3.81	&3.12&2.43&3.28&2.77&	4.92 \\
$^{276}$110&12.42&9.81	&9.66&10.87& 10.73 & 10.6	&-6.71	&-6.44	&-6.27	&-6.92	&-0.50	&-0.80	&-0.25	&-0.70	&-0.09&-0.43&0.15&-0.29&	-2.96 \\
$^{278}$110&9.38 &9.22	&9.22&10.23& 10.41 & 10.0	&0.73	&0.32	&0.95	&0.50	&1.25	&0.79	&1.45	&1.01	&1.23&0.78&1.43&1.00&	-2.13 \\
$^{280}$110&8.97 &8.62	&8.69&9.42& 9.05	 & 9.3	&2.04	&1.51	&2.22	&1.77	&3.21	&2.57	&3.36	&2.94	&2.95&2.34&3.10&2.68&	1.78 \\
$^{282}$110&8.39 &8.04	&8.31&8.52& 7.76	 &		&4.00	&3.29	&4.12	&3.69	&5.31	&4.48	&5.39	&5.01	&4.31&3.57&4.41&4.00&	6.41 \\
$^{284}$110&7.87 &7.67	&8.05&7.87& 7.57	 &		&5.96	&5.07	&6.02	&5.62	&6.79	&5.82	&6.82	&6.45	&5.29&4.46&5.37&4.94&	7.18 \\
$^{272}$108&10.24&10.26	&10.10&9.58& 9.2	 & 10.1	&-2.33	&-2.43	&-2.04	&-2.53	&-2.38	&-2.48	&-2.09	&-2.58	&-1.95&-2.09&-1.68&-2.15&	0.61 \\
$^{274}$108&9.18 &9.50	&9.01&9.52& 9.46	 & 9.5  &0.68	&0.31	&0.87	&0.44	&-0.28	&-0.56	&-0.06	&-0.51	&1.21&0.81&1.39&0.98&	-0.18 \\
$^{276}$108&8.35 &8.37	&8.35&9.06& 9.32	 & 8.8	&3.40	&2.81	&3.51	&3.13	&3.34	&2.75	&3.46	&3.08	&3.43&2.83&3.54&3.16&	0.24 \\
$^{278}$108&8.11 &7.80	&7.89&8.76& 8.53	 &		&4.27	&3.60	&4.36	&3.97	&5.47	&4.70	&5.52	&5.17	&5.10&4.36&5.16&4.80&	2.78 \\
$^{280}$108&7.74 &7.37	&7.52&8.01& 7.56	 &		&5.71	&4.92	&5.76	&5.37	&7.22	&6.31	&7.23	&6.89	&6.57&5.72&6.60&6.24&	6.43 \\
$^{268}$106&7.82 &7.78	&7.86&7.94& 7.59	 & 8.4	&4.58	&3.97	&4.65	&4.41	&4.77	&4.15	&4.83	&4.60	&4.46&3.86&4.53&4.28&	5.49 \\
$^{270}$106&8.71 &8.59	&8.74&8.16& 8.16	 & 9.1	&1.46	&1.09	&1.61	&1.24	&1.86	&1.46	&2.00	&1.64	&1.35&0.99&1.50&1.13&	3.33 \\
$^{272}$106&8.35 &8.52	&8.19&7.8& 8.39  & 8.3	&2.66	&2.20	&2.78	&2.41	&2.08	&1.66	&2.21	&1.82	&3.22&2.72&3.33&2.97&	2.54 \\
$^{274}$106&7.46 &7.63	&7.47&7.44& 8.31	 & 		&6.03	&5.31	&6.05	&5.75	&5.34	&4.67	&5.38	&5.05	&6.00&5.28&6.03&5.72&	2.8 \\
$^{276}$106&6.96 &6.81	&6.96&6.84& 7.2	 &		&8.18	&7.29	&8.14	&7.87	&8.91	&7.97	&8.85	&8.60	&8.17&7.29&8.14&7.86&	7.13 \\
$^{264}$104&6.82 &6.71	&7.02&7.87& 7.32	 & 8.14	&7.96	&7.21	&7.92	&7.82	&8.49	&7.71	&8.44	&8.35	&7.07&6.38&7.06&6.93&	5.76 \\
$^{266}$104&6.67 &6.51	&6.83&7.33& 6.73	 & 7.5	&8.66	&7.86	&8.60	&8.48	&9.46	&8.60	&9.38	&9.28	&7.93&7.18&7.90&7.75&	8.38 \\
$^{268}$104&7.47 &7.34	&7.53&7.79& 7.16	 & 8.1	&5.18	&4.62	&5.22	&4.95	&5.71	&5.12	&5.73	&5.49	&4.92&4.38&4.96&4.69&	6.46 \\
$^{270}$104&7.39 &7.39	&7.18&7.48& 7.29	 &		&5.48	&4.90	&5.51	&5.21	&5.50	&4.92	&5.53	&5.24	&6.36&5.72&6.36&6.10&	5.9 \\
$^{272}$104&6.82 &6.86	&7.64&7.21& 7.22	 &		&7.96	&7.21	&7.92	&7.67	&7.77	&7.03	&7.73	&7.48	&4.50&3.99&4.56&4.20&	6.2 \\
$^{260}$102&6.35 &6.10	&6.40&7.47& 7.24	 & 7.7	&9.35	&8.64	&9.27	&9.24	&10.64	&9.85	&10.52	&10.53	&9.11&8.41&9.03&9.00&	5.29 \\
$^{262}$102&5.87 &5.77	&6.13&6.86& 6.59	 & 7.1	&11.88	&11.02	&11.73	&11.75	&12.47	&11.56	&12.30	&12.33	&10.48&9.69&10.36&10.33&	8.18 \\
$^{264}$102&5.71 &5.52	&5.94&6.37& 6.0	 & 6.58	&12.82	&11.90	&12.65	&12.65	&13.99	&12.99	&13.78	&13.82	&11.50&10.65&11.35&11.32&	11.16 \\ 
$^{266}$102&6.23 &6.12	&6.44&6.72& 6.31	 &		&9.94	&9.19	&9.84	&9.72	&10.51	&9.73	&10.40	&10.29	&8.88&8.20&8.81&8.65&	9.59 \\
$^{268}$102&6.17 &6.12	&8.51&6.27& 6.34	 &		&10.27	&9.50	&10.16	&10.01	&10.52	&9.74	&10.41	&10.27	&0.65&0.47&0.79&0.35&	9.4 \\
\botrule
\hline
\end{tabular}
\label{tab2}}
\end{table}
\end{landscape}
\end{document}